\documentclass[fleqn,usenatbib,hyphens]{mnras}

\usepackage{newtxtext,newtxmath}
\usepackage[T1]{fontenc}

\DeclareRobustCommand{\VAN}[3]{#2}
\let\VANthebibliography\thebibliography
\def\thebibliography{\DeclareRobustCommand{\VAN}[3]{##3}\VANthebibliography}

\usepackage{graphicx}	% Including figure files
\usepackage{amsmath}	% Advanced maths commands
\usepackage{orcidlink} 
\usepackage{comment}

\newcommand{\Fiducial}{\textit{Fiducial}}
\newcommand{\FG}{\textit{FG09}}
\newcommand{\Shield}{\textit{Shield}}
\newcommand{\NoSF}{\textit{NoSF}}
\newcommand{\mbf}[1]{\mbox{\boldmath ${#1}$}}

\newcommand{\asinh}{\sinh^{-1}}

\title[DM mass constraint with astrophysical effects]{Impact of astrophysical effects on the dark matter mass constraint with 21cm intensity mapping}

\author[K. Murakami et al.]{
Koya Murakami,$^{1}$\thanks{E-mail: koya.murakami9627@gmail.com}
Atsushi J. Nishizawa,$^{2,3,4}$\,\orcidlink{0000-0002-6109-2397}\thanks{atsushi.nishizawa@iar.nagoya-u.ac.jp}
Kentaro Nagamine,$^{5,6,7}$\,\orcidlink{0000-0001-7457-8487},
and Ikko Shimizu$^8$
\\
$^1$Department of Physics, Nagoya University, Furocho, Chikusa, Nagoya, 464-8602, Japan\\
$^2$DX Center, Gifu Shotoku Gakuen University, Takakuwa-Nishi, Yanaizu, Gifu, 501-6194, Japan\\
$^3$Institute for Advanced Research, Nagoya University, Furocho, Chikusa, Nagoya, Aichi, 464-8602, Japan\\
$^4$Kobayashi Maskawa Institute, Nagoya University, Furocho, Chikusa, Nagoya, Aichi, 464-8602, Japan\\
$^5$Theoretical Astrophysics, Department of Earth and Space Science, Graduate School of Science, Osaka University, Toyonaka, Osaka 560-0043, Japan\\
$^6$Kavli IPMU (WPI), The University of Tokyo, 5-1-5 Kashiwanoha, Kashiwa, Chiba, 277-8583, Japan\\
$^7$Department of Physics \& Astronomy, University of Nevada, Las Vegas, 4505 S. Maryland Pkwy, Las Vegas, NV 89154-4002, USA\\
$^8$Shikoku Gakuin University, 3-2-1 Bunkyocho, Zentsuji, Kagawa, 765-8505, Japan
}

\date{Accepted XXX. Received YYY; in original form ZZZ}

\pubyear{2023}

\begin{document}
\label{firstpage}
\pagerange{\pageref{firstpage}--\pageref{lastpage}}
\maketitle

\begin{abstract}
We present an innovative approach to constraining the non-cold dark matter model using a convolutional neural network (CNN). 
We perform a suite of hydrodynamic simulations with varying dark matter particle masses and generate mock 21cm radio intensity maps to trace the dark matter distribution at $z = 3$ in the post-reionization epoch.
Our proposed method complements the traditional power spectrum analysis. We compare the results of the CNN classification between the mock maps with different dark matter masses
with those from the 2D power spectrum of the differential brightness temperature map of 21cm radiation. We find that the CNN outperforms the 
power spectrum.
Moreover, we investigate the impact of baryonic physics on the dark matter model constraint, including star formation, self-shielding of HI gas, and UV background model. We find that these effects may introduce some contamination in the dark matter constraint, but they are insignificant compared to the 
system noise of the SKA instruments.
\end{abstract}

\begin{keywords}
Cosmology -- dark matter -- data analysis
\end{keywords}

%%%%%%%%%%%%%%%%%%%%%%%%%%%%%%%%%%%%%%%%%%%%%%%%%%

%%%%%%%%%%%%%%%%% BODY OF PAPER %%%%%%%%%%%%%%%%%%

\section{Introduction}\label{sec:intro}

The $\Lambda$CDM model is currently the widely accepted cosmological model.
It assumes that dark matter (DM) is cold, meaning that its particle mass ($m_{\rm DM}$) is heavy enough that dark matter particles were non-relativistic at the time of freeze-out. The $\Lambda$CDM model does not make any concrete assumptions about $m_{\rm DM}$, but 
this parameter is crucial for determining the correct dark matter model. For example, sterile neutrino dark matter models predict a range of $m_{\rm DM}$ ranges from 1\,keV to 1\,MeV  \citep{Boyarsky2019}, while the weakly interacting massive particle (WIMP) model predicts a range of $m_{\rm DM}$ from 10\,GeV to 1\,TeV \citep{Alvarez2020}.

The dark matter particle mass impacts the distribution of dark matter in the universe on small scales, allowing us to estimate $m_{\rm DM}$ through the analysis of the dark matter distribution. One approach is to use the power spectrum of Lyman-$\alpha$ forest, which traces the dark matter distribution.  Previous studies have shown that the dark matter particle mass must be heavier than $\mathcal{O}(1)\  \mathrm{keV}$ \citep{Viel2013, Garzilli2019,Garzilli2019a, 2023PhRvD.108b3502V}. However, this constraint is insufficient to differentiate between different dark matter models. Therefore, developing new methods capable of extracting more information from the dark matter distribution is essential. 

This paper aims to constrain the mass of dark matter by analyzing the matter distribution in the universe using a neural network (NN). NNs are a machine learning (ML) algorithm used for big-data analysis.
They can extract information from labelled data without requiring humans to decide which data features to use. There are various kinds of NNs, and here we focus on using Convolutional Neural Networks (CNN) to extract information from images.
For example, CNNs are commonly used to distinguish between images of dogs and cats or detect human faces in images with exceptionally high accuracy.

NNs have also proven to be valuable tools in cosmology. Traditional analytical techniques, such as the two-point correlation of the matter-density distribution, can only obtain a limited amount of information from the observed data. In contrast, an ML algorithm can extract complex information from the data and capture various essential features. For instance, CNNs have been used to constrain cosmological parameters in the fields of weak lensing cosmology \citep{Ribli2019a}, simulated convergence maps \citep{Ribli2019}, and the large-scale structure \citep{Pan2019}. CNN is also applied to constrain the mass of dark matter; for example, \cite{2023arXiv230414432R} uses CNN to infer the mass of warm dark matter for N-body dark matter simulations.
Other examples include using U-net to detect signals of the Sunyaev-Zel'dovich effect by first extracting feature and then applying up-convolution to retain the original image resolution \citep{Bonjean2020}, distinguishing modified gravity models from the standard model using  CNNs \citep{Peel2019} and using NNs to reconstruct the initial conditions of the universe from galaxy positions and luminosity data \citep{Modi2018}.
These previous studies have shown that NNs often outperform traditional analytical techniques. 
 
Although dark matter cannot be observed directly, many observables, such as Lyman-$\alpha$, galaxy clustering, and weak lensing, can trace its distribution. This work focuses on the intensity mapping of the 21cm radiation emitted from neutral hydrogen (HI) due to hyperfine splitting.
Numerous ongoing or planned observations for the 21cm radiation include the Murchison Wide-field Array (MWA) \citep{MWA}, Canadian Hydrogen Intensity Mapping Experiment (CHIME) \citep{CHIME}, Hydrogen Intensity and Real-time Analysis eXperiment (HIRAX) \citep{HIRAX}, and Square Kilometer Array (SKA) \citep{SKA}.
These surveys will provide us with the HI distribution, which we can use to trace the distribution of dark matter. 

This work focuses on the HI at $z=3$, where most of the HI in the universe is ionized, and only a tiny fraction of residuals lay within the halo. During the reionization epoch, HI distribution is affected by ionisation processes, which is highly uncertain due to the complex astrophysical effects. Therefore, we focus on the post-reionization epoch, where the HI follows the dark matter halo distribution well.

\cite{2015JCAP...07..047C} forecasts the constraint on the thermally produced warm dark matter mass using the HI power spectrum from the SKA observation data at redshift $z=$ 3, 4, and 5, where they consider the HI distribution pasted on dark matter halos in the N-body simulations. \cite{2021MNRAS.500.3162B} also uses the HI power spectrum of the 21cm intensity mapping, which is modeled based on the N-body simulation, and assumes the HI halo model, and forecasts the improvement of the constraints on the axion dark matter mass compared to the limits from the Lyman-$\alpha$ forest at $z < 3$. \cite{2024MNRAS.527..739R} investigates the CNN to constrain the warm dark matter mass based on the N-body simulations, and shows the field-level inference can outperform the power spectrum analysis.  

This paper demonstrates the potential of CNNs to constrain the mass of dark matter particles compared to the traditional two-point statistics for the data from hydrodynamic simulations. While the power spectrum of dark matter or HI distribution can only extract the information of the two-point statistics, CNNs can also utilize additional information from images of the matter distribution. 
We conduct hydrodynamic simulations for CDM and NCDM models with various dark matter masses. Subsequently, we compare CNN's model discrimination ability with that of the power spectrum.

Furthermore, we take into account two practical effects that exist in observations. Firstly, we use images with varying astrophysical assumptions, such as the effects of self-shielding of HI gas, star formation, and UV background model.  We refer to these assumptions collectively as the `astrophysical model' in the following. These models can influence the ionization of HI gas, potentially affecting the results of our analysis. For instance,  a previous study \citep{Villanueva-Domingo_2021} removed the astrophysical effects from the 21cm map and used machine learning to create a map of the underlying matter density field.  
In contrast, our study uses the 21cm map with the effects of the astrophysical model and directly constrains the dark matter particle mass.
Secondly, we use images that include the system noise of SKA observations. This noise can contaminate the power spectrum calculated from observational data on small scales, affecting our analysis. 

Note that we use the projected 2-dimensional distribution of HI assuming that we stack the multiple frequency bands instead of its 3-dimensional distribution. 
This is because the signal-to-noise ratio of the single frequency bands is not enough for our assumed dataset. In other words, the thin slice of the density contrast is mostly dominated by the shot-noise.
We ignore the redshift space distortion and
the light-cone effects by the projection.

This paper is structured as follows. In Section \ref{sec:simulation}, we introduce non-cold dark matter models and discuss the relationship between the HI distribution and the differential brightness temperature. We then describe our simulation suite and the construction of our training and validation datasets. 
In Section \ref{sec:method}, we show the calculation of the power spectrum, our CNN architecture, and the procedure performed by our CNN. In Sections \ref{sec:result} and \ref{sec:summary}, we present and discuss the results of our CNN analysis and summarize our work.

Throughout this paper, we use the cosmological parameters taken from Planck 2018 \citep{Akrami2018}, except for the dark matter particle's mass.

%======================================================================================
\section{Simulations and Initial Conditions}\label{sec:simulation}
%======================================================================================
%--------------------------------------------------------------------------------------
\subsection{Non-cold Dark Matter Model}\label{ssec:wdm}
%--------------------------------------------------------------------------------------

\begin{figure}
    \centering
    \includegraphics[width=\linewidth]{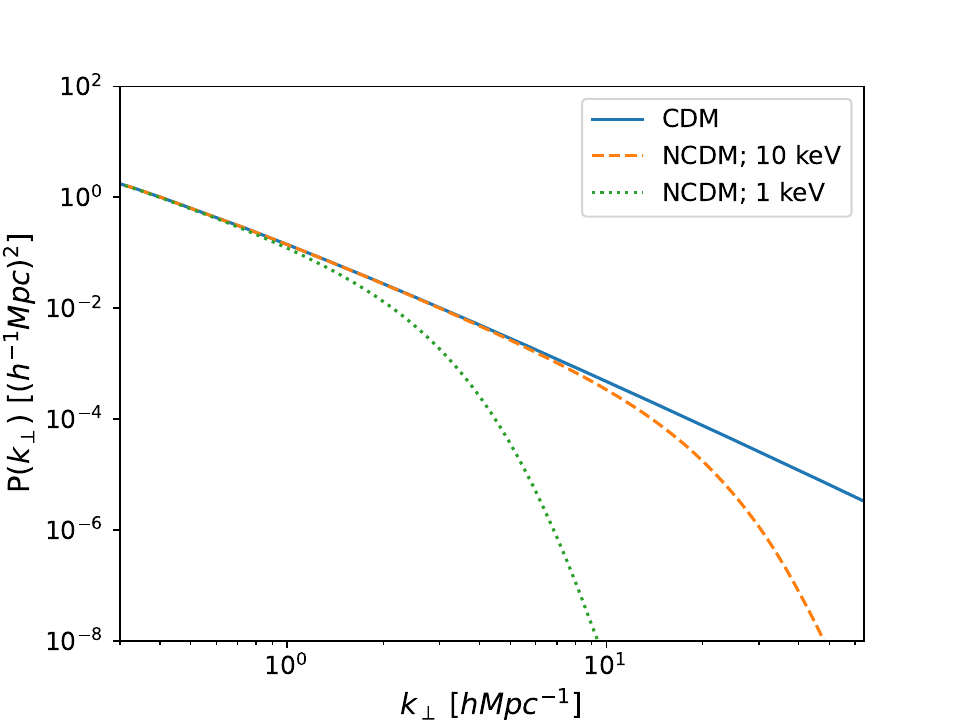}
    \caption{Examples of the theoretical linear 2D matter power spectra at $z=3$ calculated by projecting the 3D linear power spectrum along the line of sight with 50 $h^{-1}$Mpc width.
    }
    \label{fig:ini_powspec}
\end{figure}

Dark matter has a non-zero mass but interacts with electromagnetic radiation very weakly, if at all. Consequently, dark matter cannot be observed directly and can only be detected by its gravitational interactions. The gravity of dark matter influences the structure formation in the universe, so observing the large-scale structure of the universe provides information about dark matter. 

This paper considers two types of dark matter: cold dark matter (CDM) and non-cold dark matter (NCDM). 
CDM is a heavy particle that was non-relativistic at the time of freeze-out, resulting in a negligible velocity dispersion. 
In contrast, NCDM is a lighter particle with a significant velocity dispersion.

Dark matter's velocity dispersion impedes the growth of structure, particularly on small scales.  The velocity dispersion is inversely proportional to the dark matter particle's mass, $m_{\rm DM}$.  
Consequently, the damping scale of the matter power spectrum resulting from this velocity dispersion is also $\propto 1 / m_{\rm DM}$ \citep{Boyanovsky2011}. 
Fig.~\ref{fig:ini_powspec} shows the linear 2D matter power spectrum at $z=3$, where the matter distribution is projected on a 2D plane over a thickness of 50\,$h^{-1}\,\mathrm{Mpc}$ along the line of sight. We can see the suppression of the amplitude of the power spectrum by the free streaming of NCDM.
We calculate the matter power spectra for both the CDM  and NCDM models with different particle masses using the Cosmic Linear Anisotropy Solving System (\texttt{CLASS})  \citep{Lesgourgues2011a}.
Our NCDM model considers the sterile neutrino, a fundamental particle added to the standard model and distinct from active neutrinos (electron, mu, and tau neutrinos). \texttt{CLASS} uses the energy distribution function of dark matter based on the widely studied sterile neutrino model \citep{Dodelson1994} and calculates the time evolution of density perturbations, fluid-velocity divergence, and shear stress in the phase space using the fluid approximation \citep[Sec 3 of ][]{Lesgourgues2011b}. 

In this work, in addition to the CDM model, we consider the NCDM model with particle masses $m_{\rm DM}$
uniformly sampled in logarithmic scale from $1$ to $100$ keV. 
We do not consider the case of $100$ keV as it is indistinguishable from the CDM model using any of the methods described in this paper.

%--------------------------------------------------------------------------------------
\subsection{HI Distribution and Differential Brightness Temperature}\label{ssec:hi}
%--------------------------------------------------------------------------------------
This work focuses on the HI gas distribution as a tracer of the dark matter distribution. This subsection demonstrates the relationship between the HI density and the brightness differential temperature $\delta T_b$, which is the observable quantity.
We consider only the epoch well after reionization, during which most hydrogen has already been ionized.

$\delta T_b$ is the difference between the temperatures of the 21cm radiation and the 
cosmic microwave background, $T_{\gamma}$
\citep{Field1958},
\begin{equation} \label{dTb_def}
    \delta T_b = \frac{T_{\rm S} - T_{\gamma}(z)}{1+z}(1 - e^{- \tau_{\nu_0}}),
\end{equation}
where $\nu_0 = 1420$ MHz is the frequency of 21cm radiation at the rest frame, $T_{\rm S}$ is the spin temperature of HI, and $z$ is the redshift of the source of the 21cm radiation. The optical depth of HI can be given by,
\begin{equation} \label{tau0}
    \tau_{\nu_0} = \frac{3}{32 \pi} \frac{h_p c^3 A_{10}}{k_{\rm B} T_S \nu^2_0} \frac{n_{\rm HI}}{(1+z)(d v_{\parallel} / d r_{\parallel})},
\end{equation}
where $n_{\rm HI}$ is the HI number density, 
and $d v_{\parallel} / d r_{\parallel}$ is the velocity gradient of the HI gas along %with 
the line of sight. 
We replace it with the $H(z)$ because the peculiar velocity is small enough compared to the Hubble expansion \citep{Ando2021}.
We can also assume $\tau\ll 1$, and thus we have
\begin{equation} \label{eq:dTb}
    \delta T_b \sim \frac{3}{32 \pi} \frac{h_p c^3 A_{10}}{k_{\rm B} \nu^2_0} \left( 1 - \frac{T_\gamma (z)}{T_{\rm S}} \right) \frac{n_{\rm HI}}{(1+z)H(z)}.
\end{equation}

The spin temperature is \citep{Field1958}
\begin{equation}
    T^{-1}_{\rm S} = \frac{T^{-1}_\gamma + x_\alpha T^{-1}_\alpha + x_c T^{-1}_{\rm K}}{1 + x_\alpha + x_c},
\end{equation}
where $T_\alpha$ and $x_\alpha$ is the temperature of Ly-$\alpha$ and its coupling coefficient, and $T_{\rm K}$ and $x_c$ is the kinetic gas temperature and its coefficient. We calculate these values following \citep{Furlanetto2006, Endo2020}.

%--------------------------------------------------------------------------------------
\subsection{Implementation to Hydrodynamic Simulation}\label{ssec:sim}
%--------------------------------------------------------------------------------------
We perform a series of hydrodynamic simulations for dark matter models with different particle masses.
For the range of $m_{\rm DM}$ under consideration, all dark matter particles only interact gravitationally after the initial condition is generated at redshift $z=99$.  
Features of dark matter models are encoded in the matter power spectrum at the initial condition. We use the cosmological parameters obtained by Planck \citep{Akrami2018} as $\Omega_m=0.311,\ \Omega_\Lambda=0.689,\ \Omega_b=0.049,\  h=0.677,$ and $\ln 10^{10} A_s=3.047$ in the CDM model. 
In addition to the standard CDM model, we 
consider NCDM (non-CDM) models with six different particle masses logarithmically sampled from $10^{3}$ to $10^{4.66}$ eV. We only consider a single dark matter component in each case. 

The matter power spectrum for the initial condition of the hydrodynamic simulation is calculated by \texttt{CLASS} \citep{Lesgourgues2011a}, as shown in Fig. \ref{fig:ini_powspec}. Using these input power spectra, we generate the initial conditions with \texttt{2LPTic} \citep{Crocce2006}, followed by applying glass realization to remove the grid pattern in the particle distribution. While the value of AUC (introduced in Section\,\ref{ssec:AUC}) increases slightly by $\sim \mathcal{O}(0.01)$ with the grid realization, it produces unrealistic features in the matter distribution for  NCDM simulations \citep{Gotz2002, Gotz2003}.

We use \texttt{GADGET3-Osaka} \citep{10.1093/mnras/stw3061,10.1093/mnras/stz098} to solve the evolution of the matter distribution.  It is a cosmological smoothed particle hydrodynamics (SPH) code based on  \texttt{GADGET-3} \citep[initially described in][]{Springel2005}, which we modified.  Our simulations use a comoving box size of 100 $h^{-1}$Mpc on a side, with $512^3$ dark matter and $512^3$ gas particles.
We generate the initial conditions at $z=99$ and terminate the simulation at $z=3$. Throughout this work, we use the simulation snapshot at $z=3$.
We follow \cite{Nagamine_2021} for the simulation setup except for the initial conditions generated from the NCDM power spectra.
The \texttt{GADGET3-Osaka}
includes models for star formation, supernova feedback, UV radiation background, and radiative cooling/heating.  We also include the self-shielding effect of HI gas, which is the obstruction of UV radiation by optically thick HI gas. The cooling is solved by the Grackle chemistry and cooling library \citep{grackle}. Therefore, we can use the HI distribution directly from these simulations, and we do not need to assume any empirical models to predict the HI distribution from the dark matter halo.

For the {\Fiducial} model, we adopt the star formation model used in the AGORA project \citep{Kim_2014,Kim_2016}, supernova feedback described in \citet{10.1093/mnras/stz098}, and the uniform UV radiation background \citep{Haardt_2012} without the effect of the self-shielding of HI gas. We conduct seven simulations for the {\Fiducial} model, CDM nd 6 NCDM.

We examine whether the effects of astrophysical and dark matter models are distinguishable. For this purpose,
we conduct three additional simulations for CDM with different astrophysical models where some assumptions differ from the {\Fiducial} model. The {\Shield} model 
includes the effect of
self-shielding of the HI gas, 
the {\NoSF} model ignores the effect of star formation, and the {\FG} model adopts the UV radiation background model of \citep{Faucher_Gigu_re_2009} instead of \citep{Haardt_2012}.
The details of {\Fiducial}, {\Shield}, and {\FG} models are discussed in \citet{Nagamine_2021}. 

%--------------------------------------------------------------------------------------
\subsection{Training and Test Sets}\label{ssec:train}
%--------------------------------------------------------------------------------------

\begin{figure*}
  \centering
  \includegraphics[keepaspectratio,width=\linewidth]{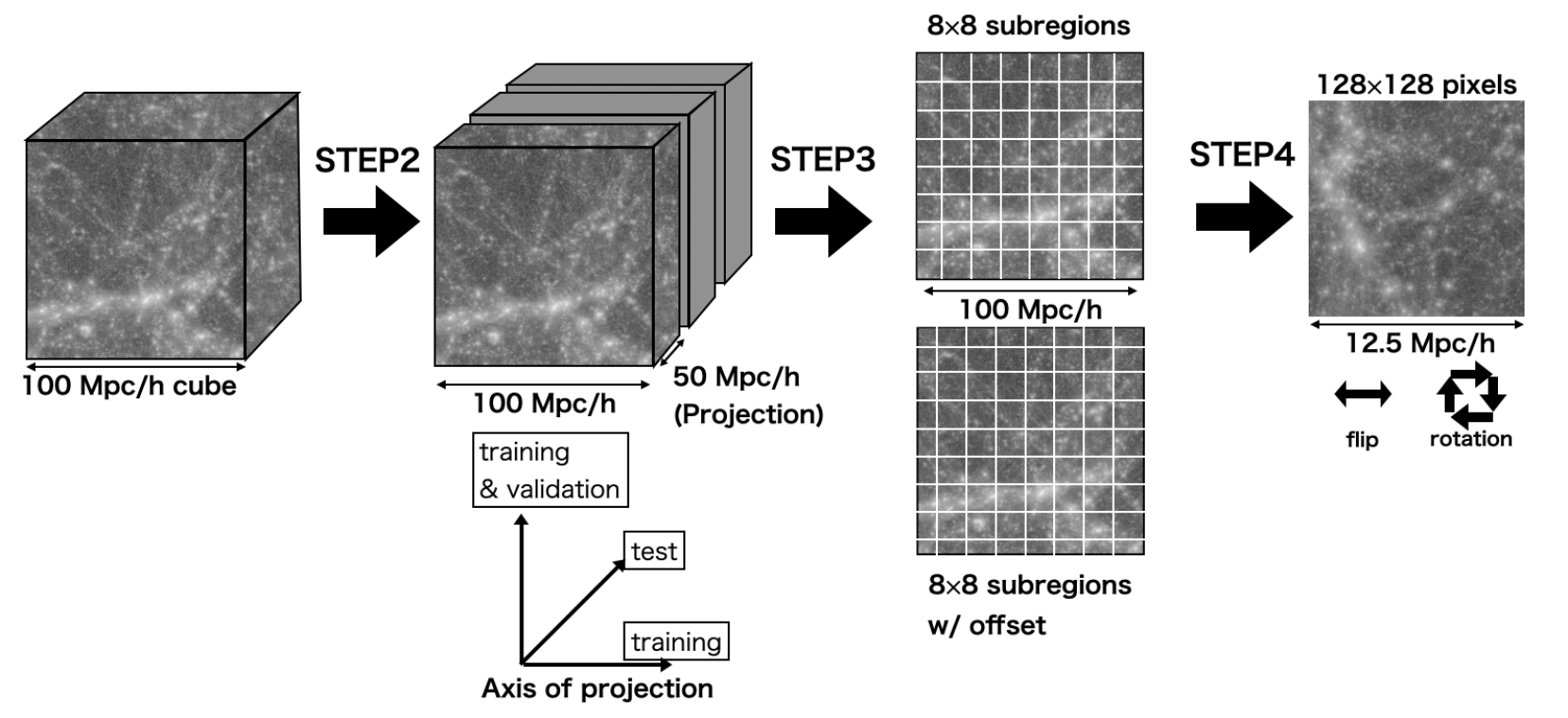}
  \caption{The procedure for making images from our simulation data.
  }
  \label{fig:gen_img}
\end{figure*}

\begin{figure}
    \centering
    \includegraphics[keepaspectratio,width=\linewidth]{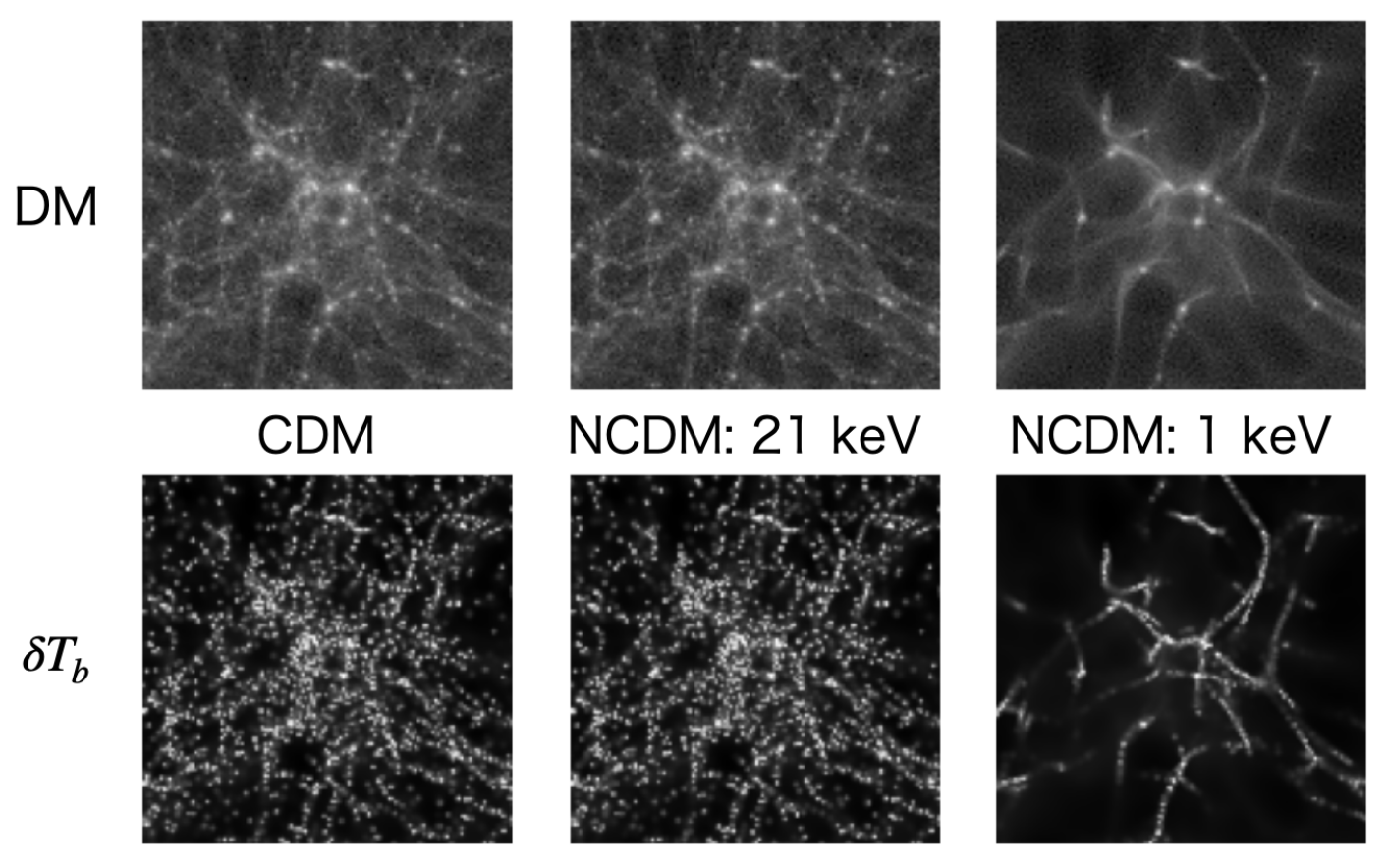}
    \caption{Example images of the CDM, 21 keV NCDM, and 1 keV NCDM models from left to right. The upper panels show dark matter images, and the lower panels show $\delta T_b$ images, which are the logarithm of the actual images for visibility. These images are made from the same region of the simulation box.}
    \label{fig:image}
\end{figure}

This subsection describes the procedures for generating the images from the hydrodynamic simulation used to train, validate, and test CNN. 
The scale of the damping of the power spectrum due to the velocity dispersion of NCDM is $k_\perp \sim 1\ h \mathrm{Mpc}^{-1}$ for $10^{3}$ eV and $k_\perp \sim 50\ h\mathrm{Mpc}^{-1}$ for $10^{4.66}$ eV in 2D Fourier space, where $k_\perp$ is the wave number perpendicular to the line-of-sight. Therefore, the image size should be sufficiently large to include the mode $k_\perp \sim 1\ h\mathrm{Mpc}^{-1}$, and at the same time, it should have sufficient resolution to resolve $k _\perp \sim 50\ h\mathrm{Mpc}^{-1}$ mode fluctuations. Our box size and the number of particles satisfy these requirements.

To generate images from a hydrodynamic simulation, we implement the following procedures (see Fig. \ref{fig:gen_img}):
\begin{itemize}
  \item{}[STEP 1] We define a $1024^3$ grid in the simulation box and redistribute the dark matter particles to the nearest grid point or calculate the HI number density in each grid using the SPH kernel. The SKA-MID system noise dominates on a smaller scale than the size of this grid. We compute the SPH kernel following the cubic spline kernel \citep{Monaghan_1985} as
  \begin{align}
    &W_{\rm SPH}(r, h) \notag \\ 
    &= A \begin{cases}
    1 - \frac{2}{3} \left( \frac{r}{h/2} \right)^2 + \frac{3}{4}\left( \frac{r}{h/2} \right)^3  & (0 < r <     \frac{h}{2}) \\
    \frac{1}{4} \left( 2 - \frac{r}{h/2} \right)^3 & (\frac{h}{2} < r < h) \\ 
    0 & (h < r) \ ,
    \end{cases}
  \end{align}
  where $h$ is the smoothing length for each particle, and $r$ is the distance between the particle and the centre of the cell. The amplitude $A$ is determined so that the sum of $W_{\rm SPH}$ overall grid becomes unity for every particle. The HI number density $n_{\rm HI}$ in a grid whose centre is located at $\mbf{x}_i$ is calculated by summing over all particles contributing to this grid,
  \begin{equation}
    n_{\rm HI} (\mbf{x}_i) = \sum_j W_{\rm SPH}(\mbf{x}_i - \mbf{x}_j | h_j)n_{\mathrm{HI},j}
  \end{equation}
  where $n_{\mathrm{HI}, j}$ is the HI number density assigned to the $j$-th particle located at $\mbf{x}_j$. And then, we calculate the $\delta T_b$ by Eq.~(\ref{eq:dTb}).
  
  \item{}[STEP 2] We divide the simulation box into three slices along the line of sight, with each width being $50$ $h^{-1}$Mpc. Each piece corresponds to the region of the simulation box from 0 to 50 $h^{-1}$Mpc, from 25 to 75 $h^{-1}$Mpc, and from 50 to 100 $h^{-1}$Mpc along the line of sight. For the test sample, we do not need to increase the number of samples by augmentations; we exclude the samples projected from 25 to 75 $h^{-1} \mathrm{Mpc}$ because they are overlapped with other slices and not totally independent. The direction of the projection is perpendicular to those for training and validation sets, as illustrated in Fig. \ref{fig:gen_img}.
  We investigate the optimal length of the slice from 50 $h^{-1}$Mpc (limited by the number of images for the sufficient training of CNN) to 0.1 $h^{-1}$Mpc (determined by the size of the cell in STEP 1) and find that AUC (introduced in Section \ref{ssec:AUC}) for the classification between the CDM and 10\,keV NCDM model is maximized when we define the projection depth as 50 $h^{-1}$Mpc.
  We have three degrees of freedom for the line-of-sight direction; these can be considered independent realizations. Therefore, we have (2 line-of-sight directions) $\times$ (3 slices) $=6$ slices for the training and validation image, and (1 line-of-sight direction) $\times$ (2 slices) $=2$ slices for the test image.
  We use the images generated from the five slices as the training data and those from the other slice as the validation data for the two line-of-sight directions. Then, we use the images from the two slices of the other line of sight direction as the test data.

  \item{}[STEP 3] Within each sub-region, the mass density of dark matter $\rho_{\rm DM} (\mbf{x})$ is integrated along the line of sight and projected onto the plane perpendicular to the line of sight, i.e., $\rho_{\rm DM} (\mbf{n}) = \int \rho_{\rm DM} (\mbf{x}) dx$ where $\rho_{\rm DM} (\mbf{n})$ is the 2D mass density of dark matter at the position $\mbf{n}$ on the 2D plane. And then, we calculate the 2D density fluctuation $\delta_{\rm DM} (\mbf{n}) = (\rho_{\rm DM}(\mbf{n}) - \bar{\rho}_{\rm DM}) / \bar{\rho}_{\rm DM}$ for dark matter, where $\bar{\rho}$ is the mean $\rho_{\rm DM}(\boldsymbol{n})$ over the simulation box.
  For HI, $\delta T_b$ is summed up along with the line of sight, i.e., $\delta T_b (\mbf{n}) = \sum_{\rm los} \delta T_b (\mbf{x})$. 
  
  \item{}[STEP 4] 
    In each slice, we cut out 8$\times$8 images. Therefore, the single image has 128$^2$ pixels, 12.5\,$h^{-1}$Mpc on a side, which is sufficiently larger scale than $k\sim 1 h \mathrm{Mpc}^{-1}$.
    For data augmentation, we employ multiple offsets when we subdivide the slices to make training or validation data. The offsets are
  $\Delta = 12.5i / 16$ $h^{-1}$Mpc where $i=0,1,\cdots, 15$ both in the directions parallel or perpendicular to a side. At the edge of the slice, we apply the periodic boundary condition.
  This may increase the number of available images sufficiently and significantly help our training process converge,
  although the shifted images are not totally independent.

\end{itemize}

In total, we have $(8 \times 8)$ (cut out in STEP 4) $\times$ (5, 1, or 3 slices in Step 2) $\times$
($16^2$, $16^2$, or $1$ (no offset)) = 81,920, 16,384, or 128 images for each training, validation, or test data for one realization of the simulation. In addition, in training the CNN, 
the images are rotated every 90 degrees and flipped horizontally to generate another different set of images.
Thus, the number of training data is effectively $81920$ $\times$ ($2$ flips) $\times$ ($4$ rotation) $=655,360$; however, in testing our CNN, test and validation images are not flipped or rotated. 
The validation data are only used for evaluating the loss to avoid overfitting and are not used to optimize the parameters.

The images for training and testing are not entirely independent, which may affect the results because they are from the same realization. To confirm whether the test images from the same realization used to make training images are valid, we prepare another realization for {\Fiducial} CDM and 10\,keV NCDM model. Then, we make 128 images from each of these new realizations using the same procedure above and use them to test our CNN trained by the training dataset from the original realization. As a result, the AUC for the images from the new realizations is 0.80, consistent with the result AUC=0.78 for the test images made from the same realization as the training images (see also Section~\ref{ssec:binary_classification}).

The image of dark matter density fluctuation $\delta$ and brightness temperature $\delta T_b$ has large dynamic ranges due to the nonlinear evolution of the structure. For our neural network architecture, it is not easy to extract feature quantities from such high dynamic range images; therefore, we apply the transformation
\begin{align}
    m_{\delta}(x) &= \asinh \left[\frac{\delta(x)}{b}\right], \\
    m_T(x) &= \asinh\left[\frac{\delta T_b (x)}{b} \right], \label{eq:asinh_dTb}
\end{align}
where $b$ is a softening parameter that controls the smooth transition scale of $\sinh^{-1}(x/b)$ from linear at $(x/b)\ll 1$ to logarithm at $(x/b)\gg 1$. We set $b=1$ for dark matter, where $b$ is dimensionless, and $b=1 {\rm nK}$ for $\delta T_b$. 
If we set $b=1$ mK, it becomes more sensitive to structures in high-density regions with less structure. Then, our CNN's performance worsens; the AUC is 0.95 for $b=1$ nK while it is 0.78 for $b=1$ mK.
More discussions on how we chose the softening parameter can be found in Section \ref{ssec:noise_effect}.

This transformation is motivated by the magnitude system, \textit{Luptitude} introduced by the Sloan Digital Sky Survey \citep{Lupton+:1999}. This is particularly useful for reducing the dynamic range, including negative values to which a simple logarithmic scale cannot be applied. In Fig. \ref{fig:image}, we show the examples of the images of DM and $\delta T_b$ for the CDM model and two NCDM models.

%======================================================================================
\section{Method}\label{sec:method}
%======================================================================================
%======================================================================================
\subsection{Power Spectrum}\label{ssec:xi}
%======================================================================================
\begin{figure}
    \centering
    \includegraphics[width=\linewidth]{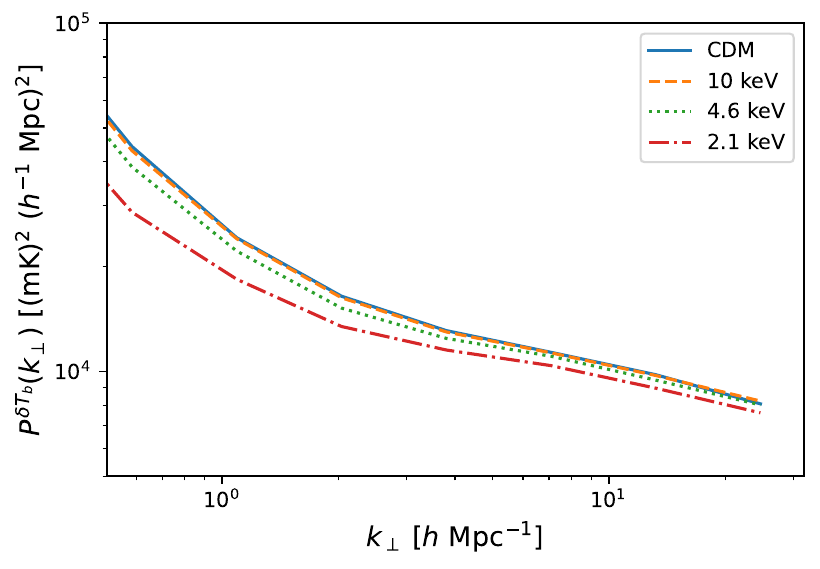}
    \caption{
    The 2D power spectra of the brightness temperature measured from the simulation with the projection length $50\,h^{-1}\mathrm{Mpc}$ for CDM (blue-solid), 10\,keV (orange-dashed), 4.6 keV (green-dotted), and 2.1 keV (red-dash-dotted) models, respectively.
   }
    \label{fig:2d_dTb_pk}
\end{figure}

In many cases of cosmological inference, the clustering analysis is mainly quantified through the two-point statistics such as power spectrum or correlation function in the literature because of the great success of the linear perturbation theory and inflation model to predict the Gaussian initial density field.  
However, the nonlinear gravitational evolution of the structure carries additional information than two-point statistics.
In this section, we revisit the basic methodology of the power spectrum-based analysis. Note that, unlike the parameter inferences in which we compare the data with the prediction, here we focus on the classification problem: whether we can distinguish the power spectra of NCDM from the {\Fiducial} power spectrum of CDM. For this purpose, we use the 2D power spectrum measured from test images generated in Section \ref{ssec:train}. In general, 3D power spectra of the 21 cm signals are used for cosmological analysis. However, our purpose is to compare CNN for the 2D images from the 21cm signals and the two-point statistics.  Therefore, here we consider the 2D power spectrum of the images for CNN usage.
Two dimensional Fourier counterpart $\tilde{A} (\mbf{k}_\perp) $ of a physical quantity $A(\mbf{n})$ defined on a two dimensional position $\mbf{n}$ is written as
\begin{equation}
      \tilde{A} (\mbf{k}_\perp) = \int \exp{(-i \mbf{k}_\perp \cdot \mbf{n})} A (\mbf{n}) d^2 n.
\end{equation}
And then, using the simulation, the power spectrum for the projected field along the line of sight for dark matter is
\begin{equation} \label{eq:pk_dm}
 P^{\rm DM}(k_{\perp, i})\ [(h^{-1} \mathrm{Mpc})^2] = \frac{1}{L^2} \frac{1}{N_{k_{\perp, i}}} \sum_j \tilde{\delta}_{\rm DM} (\mbf{k}_{\perp, j}) \tilde{\delta}_{\rm DM}^*(\mbf{k}_{\perp,j}),
\end{equation}
and the one for the $\delta T_b$
 is 
\begin{equation} \label{eq:pk_dTb}
 P^{\rm \delta T_b}(k_{\perp,i})\ [\mathrm{mK}^2 (h^{-1} \mathrm{Mpc})^2] = \frac{1}{L^2} \frac{1}{N
_{k_{\perp,i}}} \sum_j \tilde{\delta T_b} (\mbf{k}_{\perp,j}) \tilde{\delta T_b}^*(\mbf{k}_{\perp,j}),
\end{equation}
where $\tilde{\delta}_{\rm DM}$ and $\tilde{\delta T_b}$ are two dimensional Fourier counterparts of $\delta_{\rm DM}$ and $\delta T_b$ 
respectively, 
$k_{\perp,i} = |\mbf{k}_{\perp,i}|$ is the absolute value of the wave number of the center of the $i$-th 
bin, $\mbf{k}_{\perp,j}$ is 
the wave number satisfies $k_{\perp,i} \le |\mbf{k}_{\perp,j}| < k_{\perp,i+1}$, $N_{k_{\perp,i}}$ is the number of the modes in $i$-th $k_\perp$ bin, and $L$ is the size of image and is 12.5 $h^{-1}$Mpc. The factor $1/L^2$ is due to the finite integration interval in the Fourier transform ($L \rightarrow 2\pi$ if the image size is infinite).
The minimum and maximum wavenumbers are automatically determined by the size and resolution of the images from which we measure the spectrum. They are $k_{\perp,\rm min}=2\pi/12.5\ h \mathrm{Mpc}^{-1}$ and  $k_{\perp,\rm max}=2\pi/(12.5/128)\ h \mathrm{Mpc}^{-1}$, respectively.
We change the number of $k_\perp$-bins from 1 to 20, and find that four is optimal in terms of the classification performance of the power spectrum. Figure \ref{fig:2d_dTb_pk} shows the 2D power spectra of $\delta T_b$. 
Unlike the dark matter power spectrum, we see the overall suppression of the amplitude for the lighter mass of dark matter models. This can be explained as follows.
In the NCDM models, small halos are suppressed to form due to the free streaming of dark matter (see Appendix~\ref{sec:hi_halo}). Therefore, the halo bias effectively gets a higher value than the CDM case. In addition, in the post-reionization epoch, most of the HI resides only within the halo; thus, the HI bias also becomes higher. However, we have to consider the overall amplitude suppression of the HI power spectrum because $P^{\delta T_b}$ is proportional to the square of the HI density, and the NCDM model suppresses the total HI abundance due to the smaller number of dark matter halos.
Therefore, the NCDM models suppress the $\delta T_b$ power spectrum even above the scales of free streaming.
Note that \cite{2015JCAP...07..047C} assumes that the HI abundance is unchanged for different dark matter models, and thus, the behaviour of the power spectrum is different, i.e., the $\delta T_b$ power spectrum for NCDM is amplified in the previous work due to the HI bias as mentioned above while our power spectrum is suppressed. 
In \cite{2015JCAP...07..047C}, the HI is pasted on the dark matter halo based on a model, and the total HI abundance is normalized by the value from the observations. On the other hand, in our simulation, the HI abundance is fixed at the initial condition by $\Omega_m$, baryon fraction, and hydrogen fraction. Then, its evolution is affected by the NCDM model as we mentioned above. Therefore, we consider the HI abundance is one of the information for the NCDM model and do not fix the HI abundance. Our power spectra are also amplified for the NCDM model when we normalize the power spectra by the HI abundance.

The covariance matrix of the power spectrum can be measured from test images of the CDM simulation,
\begin{equation} \label{eq:cov_pk}
    C_{mn} = \frac{1}{N_{\rm img}} \sum_l \left( P_l (k_{\perp,m}) - \bar{P} (k_{\perp,m}) \right) \left( P_l (k_{\perp,n}) - \bar{P} (k_{\perp,n}) \right),
\end{equation}
where the subscript $l$ is the label of the test images, $N_{\rm img} (=128)$ is the number of the test images from CDM simulation data, and $\bar{P}(k_\perp)= \sum_l P_l (k_\perp)/N_{\rm img}$ is the mean of the power spectra of the CDM test image.

%======================================================================================
\subsection{Convolutional Neural Network}\label{ssec:dnn}
%======================================================================================
\begin{table}
  \centering
  \begin{tabular}{| c || c | c |}
    \hline
    \ & Layer & Output map size \\ \hline
    1 & Input & $128 \times 128 \times 1$ \\
    2 & $3 \times 3$ convolution & $126 \times 126 \times 32$ \\
    3 & $3 \times 3$ convolution & $124 \times 124 \times 32$ \\
    4 & $3 \times 3$ convolution & $122 \times 122 \times 64$ \\
    5 & $3 \times 3$ convolution & $120 \times 120 \times 64$ \\
    6 & $3 \times 3$ convolution & $118 \times 118 \times 128$ \\
    7 & $1 \times 1$ convolution & $118 \times 118 \times 64$ \\
    8 & $3 \times 3$ convolution & $116 \times 116 \times 128$ \\
    9 & $2 \times 2$ AveragePooling & $58 \times 58 \times 128$ \\
    10 & $3 \times 3$ convolution & $56 \times 56 \times 256$ \\
    11 & $1 \times 1$ convolution & $56 \times 56 \times 128$ \\
    12 & $3 \times 3$ convolution & $54 \times 54 \times 256$ \\
    13 & $2 \times 2$ AveragePooling & $27 \times 27 \times 256$ \\
    14 & $3 \times 3$ convolution & $25 \times 25 \times 512$ \\
    15 & $1 \times 1$ convolution & $25 \times 25 \times 256$ \\
    16 & $3 \times 3$ convolution & $23 \times 23 \times 512$ \\
    17 & $2 \times 2$ AveragePooling & $12 \times 12 \times 512$ \\
    18 & $3 \times 3$ convolution & $10 \times 10 \times 512$ \\
    19 & $1 \times 1$ convolution & $10 \times 10 \times 256$ \\
    20 & $3 \times 3$ convolution & $8 \times 8 \times 512$ \\
    21 & $1 \times 1$ convolution & $8 \times 8 \times 256$ \\
    22 & $3 \times 3$ convolution & $6 \times 6 \times 512$ \\
    23 & GlobalAveragePooling & 1 $\times$ 1 $\times$ 512 \\
    24 & FullyConnected & 2 \\ \hline
  \end{tabular}
  \caption{Our CNN architecture. The second column shows the type of layer, and the third column shows the size of the output from the layer ((height $\times$ width $\times$ channel) of the feature map). The total number of trainable parameters is 8,328,610.}
  \label{tb:CNN_architecture}
\end{table}

In this section, we describe our CNN model.
In our model, 
we apply convolution layers with $3\times 3$ kernels for deep multiple layers to extract characteristics over various scales from images.
We use the publicly available platform PyTorch\citep{pytorch} 
to construct our model.
We follow the previous work \citep{Ribli2019a} for the architecture of the neural network, except that we skip the first two Average Pooling layers in \citep{Ribli2019a} because the size of the input image is different. The architecture is summarized in Table \ref{tb:CNN_architecture}. The total number of trainable parameters in this architecture is $\sim 8\times 10^6$; therefore, $10^5$ images are required to avoid both over- and under-fitting of the data \citep{Han2015}. Consequently, we prepare $6\times 10^5$ images for each simulation. 

We try to find the optimal number of layers, summarized in Table~\ref{tb:CNN_architecture}.
If we halve the number of layers by skipping all of the 4th, 5th, 6th, 12th, 16th, and 22nd layers in Table~\ref{tb:CNN_architecture}, the losses, computed 
by Eq.\,(\ref{eq:loss}),  
gets 10 times larger
and the validation accuracy is 
$\sim 0.5$, which means 
the model prediction is random and not able to classify the inputs.
This is because this model is too simple. Conversely, if we double the number of layers by repeating each convolution layer twice with zero padding to 
keep the size of the feature map unchanged,
the loss does not decrease during the optimization.
This is because the number of trainable parameters is too large 
compared to
the size of our training dataset and the vanishing gradients may occur \citep{2015arXiv151203385H}. Again, we observe that the validation accuracy fluctuates around 0.5.

Now, we explain the detailed procedures in each layer.
In general, $n_x \times n_y$ convolution kernel translates the $N_x \times N_y$ input image into $(N_x-(n_x-s_x))\times(N_y-(n_y-s_y))$ image, when the stride is $s_x \times s_y$ and no-padding is applied. In our analysis, we always fix $s_x=s_y=1$.
The number of output feature maps depends on the number of kernels in the current layers, which, in our analysis, can vary from 1 to 512.
The kernel values are initially set randomly, but they are subject to be optimized during the training process. After each convolution layer, we add a batch-normalization layer to normalize the distribution of the input feature map, 
which increases
the training efficiency \citep{2015arXiv150203167I}.
Also, after every convolution layer, we apply an activation function of ReLU \citep{ReLU}.

In the $n_x \times n_y$ AveragePooling layer, when we set the stride to be the same as $n_x$ and $n_y$,
% the pooling size
then $N_x \times N_y$ input image is converted into an $(N_x / n_x) \times (N_y / n_y)$ image. 
In these layers, the information in the input image is compressed and simplified.
In the GlobalAveragePooling layer, the values of all pixels in each input %feature 
map are averaged. 
We find that the combination of the GlobalAveragePooling and the single FullyConnected layers shows better performance than the multiple FullyConnected layers.
Finally, the FullyConnected layer adopts the softmax activation function as the final output of the model, which is the probabilities of the input image being CDM or NCDM models, respectively.

Now, we can express the outputs of the input and predicted classes,
\begin{equation}
  \label{eq:CNNout}
  \boldsymbol{p}_{i}(\mathrm{M}) =
  \{
  p_{i}(\mathrm{CDM} |\mathrm{M}),
  p_{i}(\mathrm{NCDM} |\mathrm{M})
  \},
\end{equation}
where $p_{i}(k|\rm M)$ is the probability predicted by our CNN that the $i$-th input image is $k(=\rm CDM\ or\ NCDM)$ model and M means the true dark matter model for the $i$-th input.
This is converted from the output of the last FullyConnected layer $\boldsymbol{y}({\rm M}) = \{ y_{i}(k|{\rm M}), y_{i}(k|{\rm M}) \}$ by the softmax function;
\begin{equation} \label{eq:softmax}
  p_i(k|\mathrm{M}) = \frac{\exp{(y_i(k|\mathrm{M}))}}{\exp{(y_{i}({\rm CDM}|\mathrm{M}))} + \exp{(y_{i}({\rm NCDM}|\mathrm{M}))}}.
\end{equation}

For loss function, we adopt a typical cross-entropy
\begin{equation}
  \label{eq:loss}
  E_i(\boldsymbol{w}) = - \sum_{k} \tilde{p}_i({k|{\rm M}}) \ln{(p_i(k|{\rm M}))}.
\end{equation}
In this equation,
the ground truth $\tilde{p}_i$ 
takes 1 for correct class ($k={\rm M}$) and 0 otherwise ($k\neq{\rm M}$), and prediction $p_i$ takes continuous values between 0 and 1. The output $p_i$ is an implicit function of $\boldsymbol{w}$, which is subject to be optimized.

For optimization purposes, we use the AMSG{\footnotesize RAD}\citep{AMSGRAD}, and take the learning rate as $10^{-5}$.
The updates of the trainable parameters are computed based on the averaged value of the loss function $\bar{E}$ over the mini-batch sample, which is randomly drawn from the training dataset. 
We take eight mini-batch samples to facilitate a better training convergence. The validation sample generated as Section \ref{ssec:train} evaluates the training. The convergence condition is that the validation loss averaged over the latest five epochs converges to 1\%.

%======================================================================================
\subsection{Evaluation of Classification}\label{ssec:evaluation}
%======================================================================================

In the following, we consider the binary model classification between the images from the CDM and NCDM models.
In this subsection, we introduce the Kolmogorov-Smirnov (KS) test, used to evaluate the classification results by CNN and power spectrum. In addition, for image classification, we also use AUC to quantify the goodness of the prediction model.

%--------------------------------------------------------------------------------------

\subsubsection{Kolmogolov-Smirnov Test}

The KS test evaluates whether the underlying distribution functions for two distinct finite samples are the same \citep{KS_1, smirnov1939estimate}. Our analysis uses the KS test to discriminate the images or power spectra of CDM and NCDM models.

We use the distributions of the $\chi^2$ values of the power spectra and the outputs from our CNN. For the $i$-th test image of dark matter model M, we calculate the $\chi^2$ value of the power spectrum as
\begin{align}
    \chi^2_{{\rm PS},i} (\rm{M}) &= 
    \boldsymbol{\Delta} P_i (k_\perp|\mathrm{M}) 
    \boldsymbol{\mathrm{C}}^{-1} 
    \boldsymbol{\Delta} P_i (k_\perp|\mathrm{M}),
\end{align}
where $\boldsymbol{\Delta} P_i(k_\perp|\mathrm{M})=P_i (k_\perp|\mathrm{M}) - \bar{P} (k_\perp|\mathrm{CDM}) $ is the power spectrum difference of the $i$-th input image of the dark matter model M defined by Eq. (\ref{eq:pk_dm}) or Eq. (\ref{eq:pk_dTb}), $\bar{P}(k_\perp) $ is the power spectrum averaged over the CDM images and $\boldmath{\mathrm{C}}^{-1}$ is the 
inverse 
of the covariance matrix defined in Eq. (\ref{eq:cov_pk}).
For the case of image classification, we have defined the discriminator that quantifies the difference between two dark matter models,
\begin{equation} \label{eq:chi_ml}
    \displaystyle
    \chi^2_{\mathrm{CNN},i} (\mathrm{M}) =
    \frac{(y_i (\mathrm{M}) - \bar{y}(\mathrm{CDM}))^2}{\frac{1}{N}\displaystyle\sum_j \left( y_j (\mathrm{CDM}) - \bar{y}(\mathrm{CDM}) \right)^2},
\end{equation}
where $y_i(M)$ is the prediction of CNN
that the $i$-th input image of dark matter model M to be the NCDM model, where $M$ is either CDM or NCDM. $\bar{y}(\mathrm{CDM})$ is the average of $y_{i}({\rm CDM})$ over the CDM test images and the denominator of the right-hand side of Eq.(\ref{eq:chi_ml}) is the variance of the CNN outputs for CDM input images.

Then, we conduct the KS test for the distribution of $\chi^2_i (\mathrm{CDM})$ and $\chi^2_i(\mathrm{NCDM})$ with \texttt{stats.ks\_2samp} method in \texttt{SciPy} \citep{scipy}.
The null hypothesis of our test is that there is no significant difference between the distribution of the $\chi^2$ for the CDM images and the NCDM images.
This work uses the significance level of $p$-value = 0.01 ($\sim 2.6\sigma$).

We note that the KS test employed here only tells us whether or not there is a significant difference between the images of the two models. Thus, it cannot quantify whether the output model is correct. To further quantify this, we will introduce AUC in the next section.

\subsubsection{AUC}\label{ssec:AUC}
The area under the Receiver Operating Characteristic (ROC) curve
is used to quantify 
the ability of the CNN model to correctly predict the dark matter model.

The output of the CNN is the probability of the input image being the NCDM model. For binary classification, we need to define a specific threshold $t$ such that the CNN can recognize the input image as the NCDM model if $p_i > t$. Therefore, we can explicitly consider the four different cases: (1) True Positive (TP) if $p_{i}(\mathrm{NCDM}|\mathrm{NCDM})\ge t$, (2) True Negative (TN) if $p_{i}(\mathrm{NCDM}|\mathrm{CDM})<t$, (3) False Positive (FP) if  $p_{i}(\mathrm{NCDM}|\mathrm{CDM})\ge t$ and (4) False Negative (FN) if  $p_{i}(\mathrm{NCDM}|\mathrm{NCDM})<t$, where all four quantities are a function of $t$.

The ROC curve can now be defined as the collection of points at which parameter $t$ continuously changes from 0 to 1. More specifically, it can be expressed in a parametric manner,
\begin{equation}
  \mathrm{ROC}:
  x(t)=\frac{\rm FP}{\rm TN+FP}, \quad 
  y(t)=\frac{\rm TP}{\rm TP+FN}, 
\end{equation}
where $x$ represents the fraction of misclassified images as the NCDM model out of all CDM test images, and $y$ represents the fraction of correctly classified images as the NCDM model given all NCDM inputs.
Therefore, the area under the curve (AUC) approaches unity when the classification is efficient and complete.

%======================================================================================
\section{Results}\label{sec:result}
%======================================================================================

In this section, we show the dark matter model 
classification results between CDM and NCDM whose particle mass is $m_{\rm DM}$ by 
using image-based CNN and compare it with the 2D power spectrum-based classification.
In Section~\ref{ssec:binary_classification}, 
we show the results for dark matter density field images and compare them with the 2D power spectrum classification. We further extend the same analysis to the $\delta T_b$ field, which is the indirect probe of dark matter but a direct observable.
In Section~\ref{ssec:astro_effect}, we 
explore how the results are affected by the nuisance effects caused by the astrophysical feedback. Finally, in Section~\ref{ssec:noise_effect}, we also consider the SKA-MID instruments' system noise, which can weaken the constraints.

\subsection{Comparison between dark matter and $\delta T_b$ image}\label{ssec:binary_classification}

\begin{figure}
    \centering
    \includegraphics[width=\linewidth]{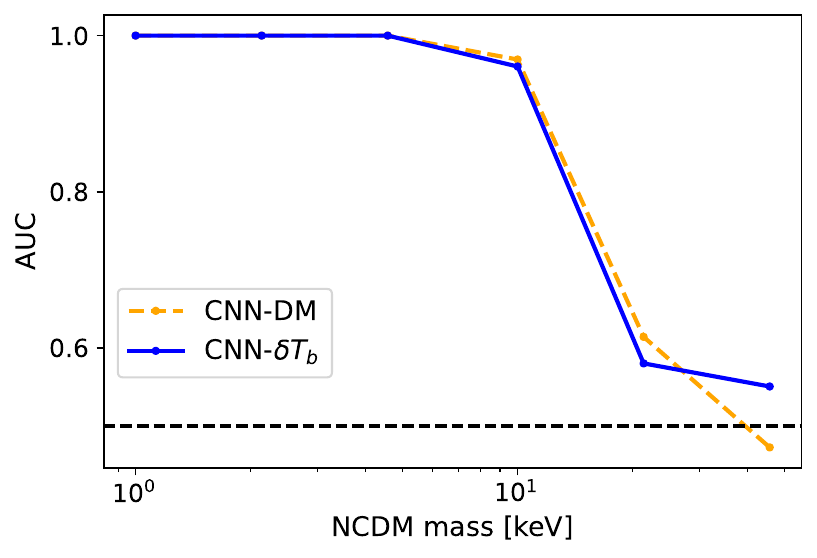}
    \caption{
    AUC as a function of $m_{\rm DM}$ for dark matter images (blue solid) and $\delta T_b$ images (orange dashed).
    The dashed horizontal line represents the case of random classification.
    We see CNN-$\delta T_b$ shows comparable performance to CNN-DM.
    }
    \label{fig:AUC_dm_dTb}
\end{figure}

\begin{figure*}
\centering
\begin{tabular}{cc}
  \includegraphics[keepaspectratio,width=0.5\linewidth]{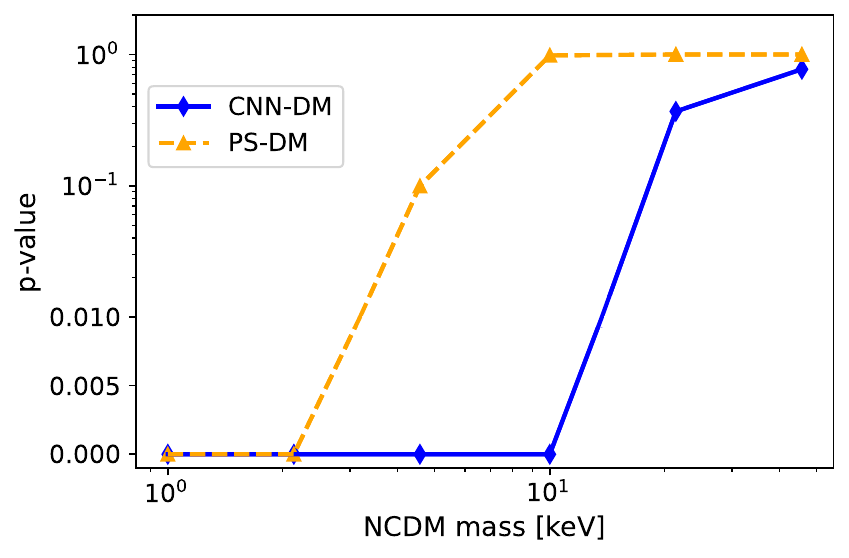} &
  \includegraphics[keepaspectratio,width=0.5\linewidth]{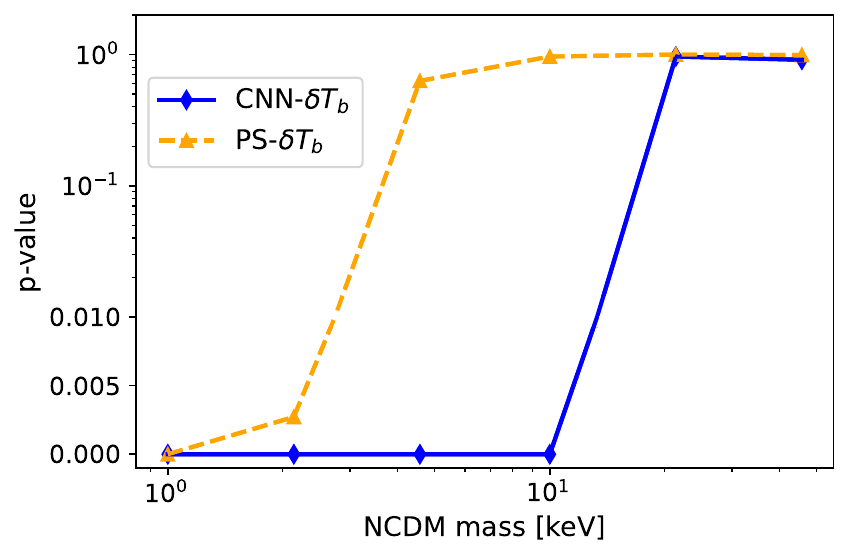}
\end{tabular}
  \caption{
  The comparison of the dark matter model classifications is evaluated by the $p$-value of the KS test. The horizontal axis is the mass of NCDM, and the vertical axis is the $p$-value for PS-DM (orange dashed) and CNN-DM (blue solid) in the left panel, and for PS-$\delta T_b$ (orange dashed) and CNN-$\delta T_b$ (blue solid) in the right panel. We see that CNN has a significantly better $p$-value than the 2D power spectrum for both the classification of dark matter and $\delta T_b$ images. 
  \label{fig:p-value_dm_dTb}
  }
\end{figure*}

For latter convenience, we first define the acronyms X-Y, denoting the observable Y is classified by the method X, where X is either CNN or PS, and Y is either DM or $\delta T_b$. E.g., CNN-$\delta T_b$ stands for the image-based classification using the $\delta T_b$ map.

First, we compare the results of CNN-DM and CNN-$\delta T_b$.
In Fig. \ref{fig:AUC_dm_dTb}, we see that for both dark matter and $\delta T_b$ images, AUCs are greater than 0.95 at mass ranges of $m_{\rm DM} \le 10$ keV. The AUC of CNN-$\delta T_b$ is comparable to the one of CNN-DM, so $\delta T_b$ is the valid tracer of the dark matter distribution for our CNN.

Next, we compare the discrimination power between CNN and PS for dark matter or $\delta T_b$ using the KS test.  Figure~\ref{fig:p-value_dm_dTb} shows the $p$-value of the KS test for classifying dark matter data on the left panel and $\delta T_b$ data on the right panel. We see that our CNN shows better performance than the 2D power spectrum for both dark matter and $\delta T_b$ data.
For example, $p$-value of CNN-DM and CNN-$\delta T_b$ at $m_{\rm DM} = 4.6$ keV is less than $0.001$ and can reject the null hypothesis with high significance while the $p$-value of PS-DM and PS-$\delta T_b$ is of the order of 0.1.

Now, we compare the results of CNN-DM and CNN-$\delta T_b$.
They show similar performance for the KS test. Both of them can classify the images for $m_{\rm DM} \le 10$ keV with high significance ($p$-value $< 0.001$, and lose the classification ability for more massive dark matter (e.g. the $p$-values of CNN-DM and CNN-$\delta T_b$ are $=$ 0.37 and $>$ 0.99 at $m_{\rm DM}=$ 21 keV). 

\subsection{Effect of Astrophysical Model}\label{ssec:astro_effect}

\begin{figure*}
    \centering
    \begin{tabular}{cc}
      \includegraphics[keepaspectratio,width=0.5\linewidth]{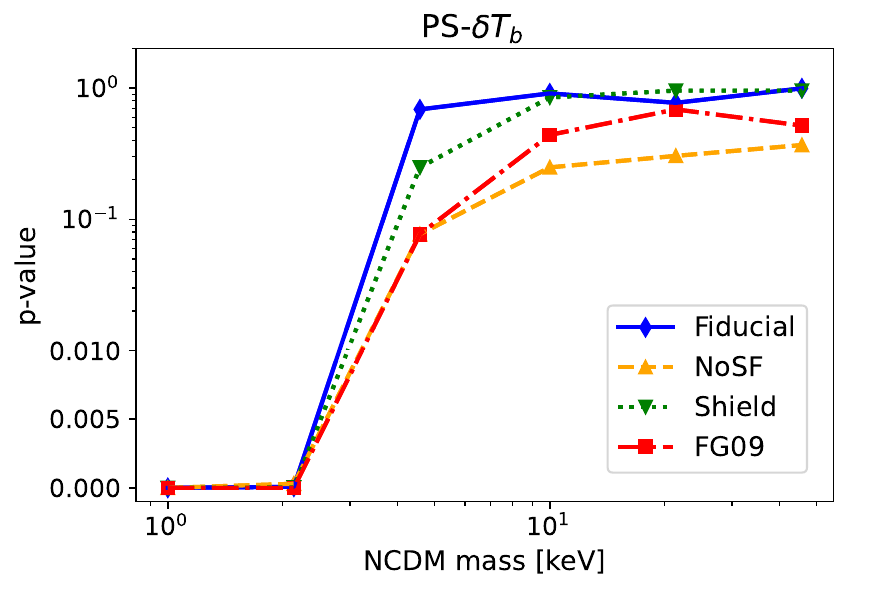} &
      \includegraphics[keepaspectratio,width=0.5\linewidth]{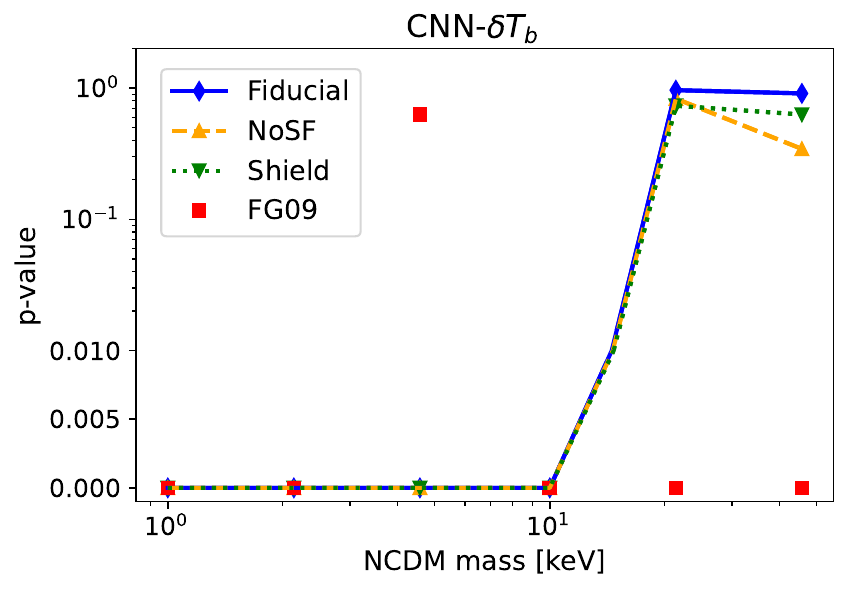}
    \end{tabular}
    \caption{
     The $p$-values of the KS test are shown, including the results for the various astrophysical models. The horizontal axis is the NCDM mass, and the vertical axis shows the $p$-value for the binary classification by PS-$\delta T_b$ (left panel) or CNN-$\delta T_b$ (right panel) between the NCDM and the CDM assumed the {\Fiducial} (blue solid), {\NoSF} (orange dashed), {\Shield} (green dotted), and {\FG}  (red square plot) model. For the {\FG} model, $p$-value significantly differs from {\Fiducial} model. The distribution of the output for {\FG} model and NCDM model is different statistically, but CNN cannot classify the images correctly, as we can see in Fig.~\ref{fig:confusion_matrix} and Table.~\ref{tb:astro_AUC}.
    }
    \label{fig:pval_astro}
\end{figure*}

\begin{figure*}
\centering
\begin{tabular}{cc}
  \includegraphics[keepaspectratio,width=\linewidth]{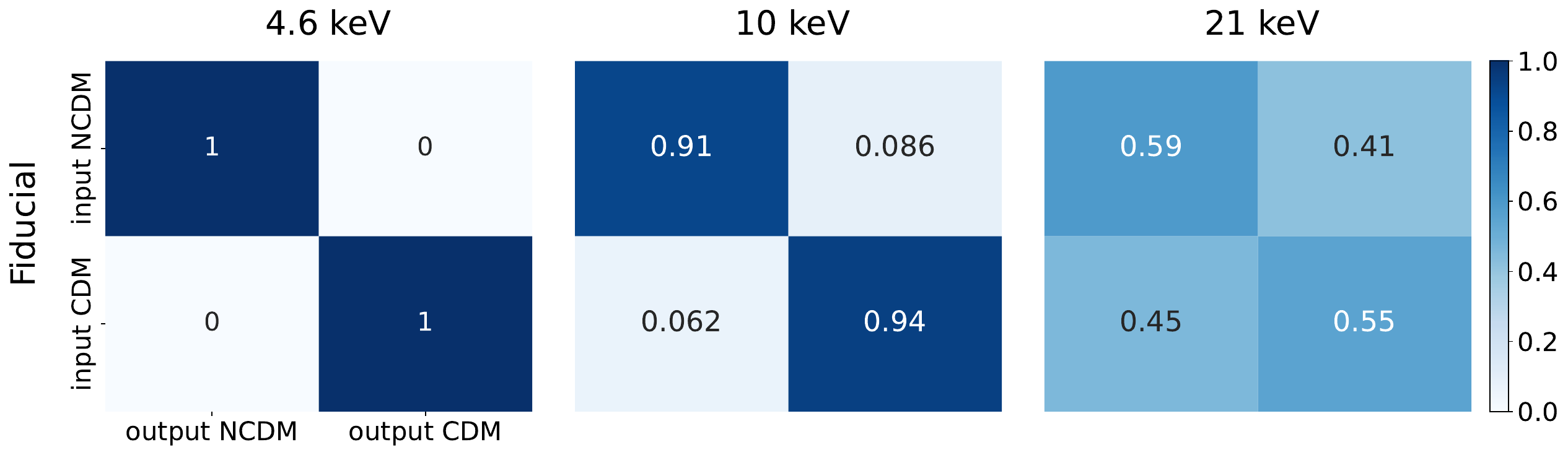} \\
  \includegraphics[keepaspectratio,width=\linewidth]{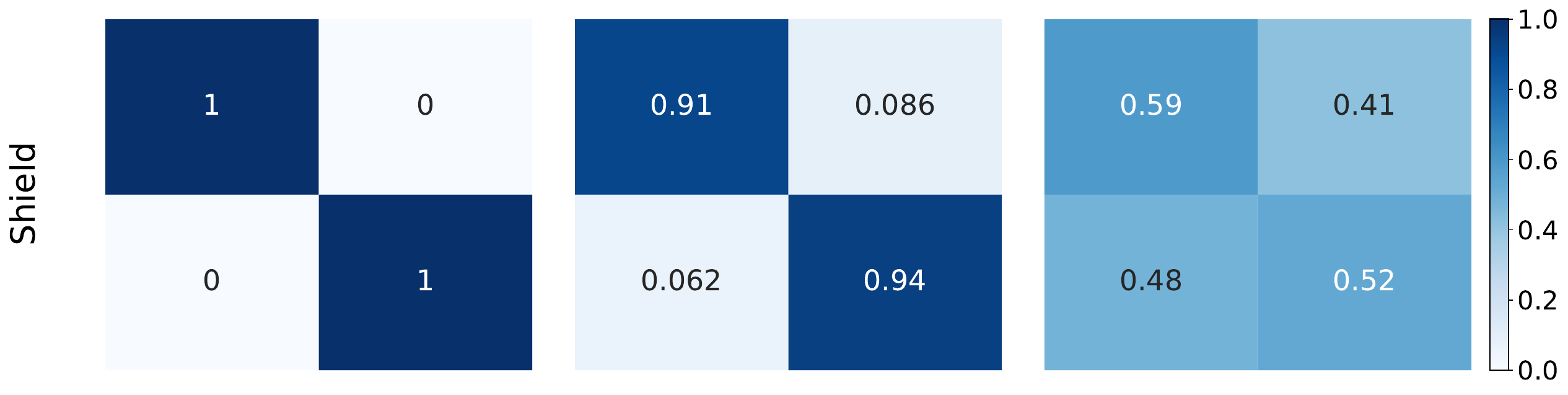} \\
  \includegraphics[keepaspectratio,width=\linewidth]{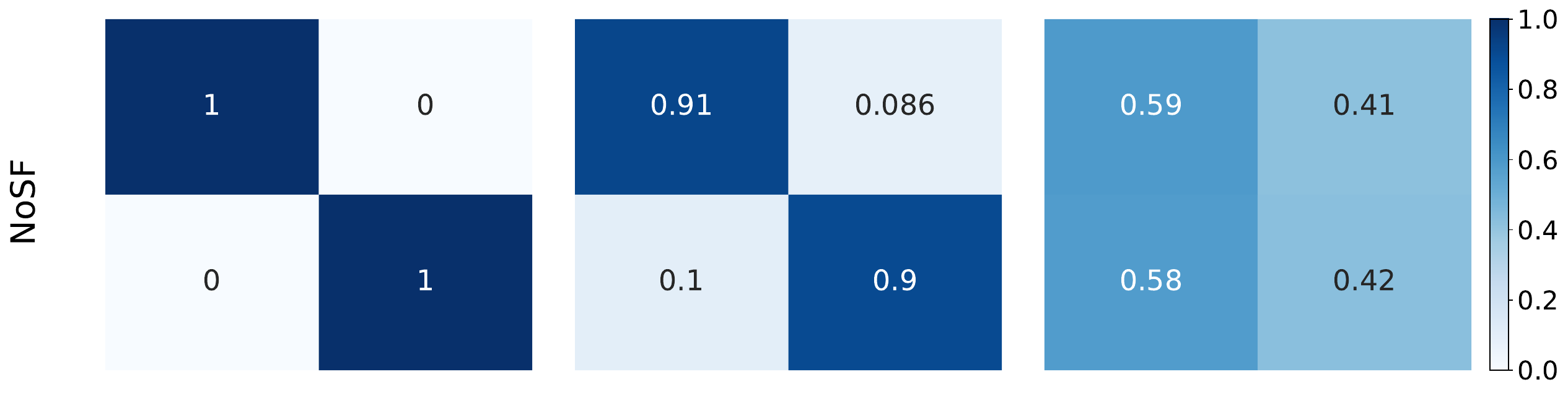} \\
  \includegraphics[keepaspectratio,width=\linewidth]{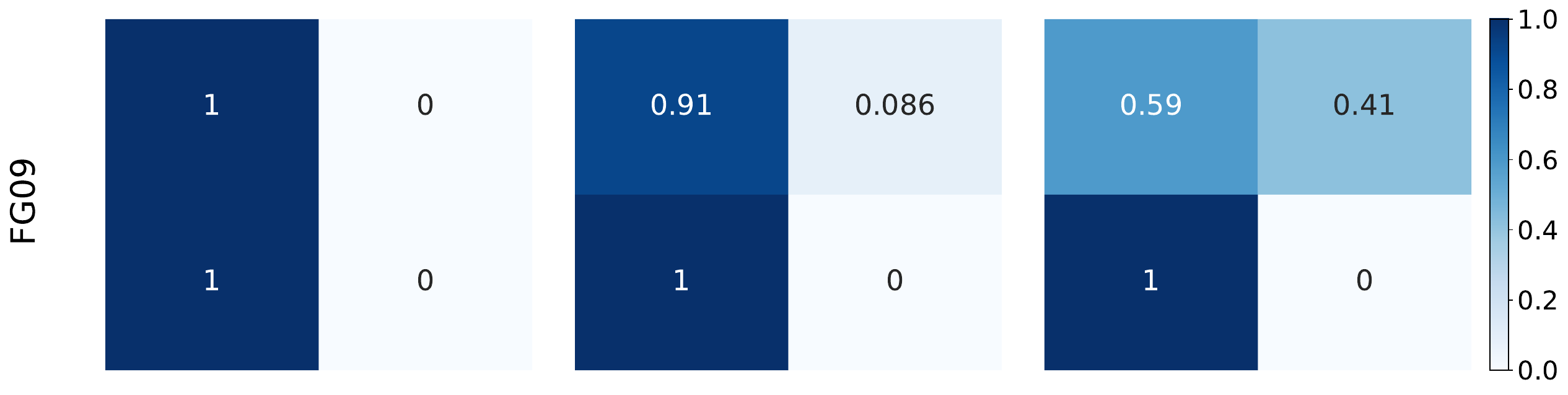}
\end{tabular}
  \caption{
  This figure shows the confusion matrix.
  Each column corresponds to the NCDM mass $m_{\rm DM} = 4.6, 10,$ and $21$\,keV from left to right, and each row corresponds to the astrophysical model: {\Fiducial}, {\Shield}, {\NoSF}, and {\FG} from top to bottom. 
  Except for the {\FG} model, the confusion matrix is almost the same as for the {\Fiducial} model. On the other hand, in the case of the {\FG} model, the CDM test images are classified into NCDM incorrectly, regardless of the mass of NCDM.
  }
  \label{fig:confusion_matrix}
\end{figure*}

\begin{figure}
    \centering
    \includegraphics[width=\linewidth]{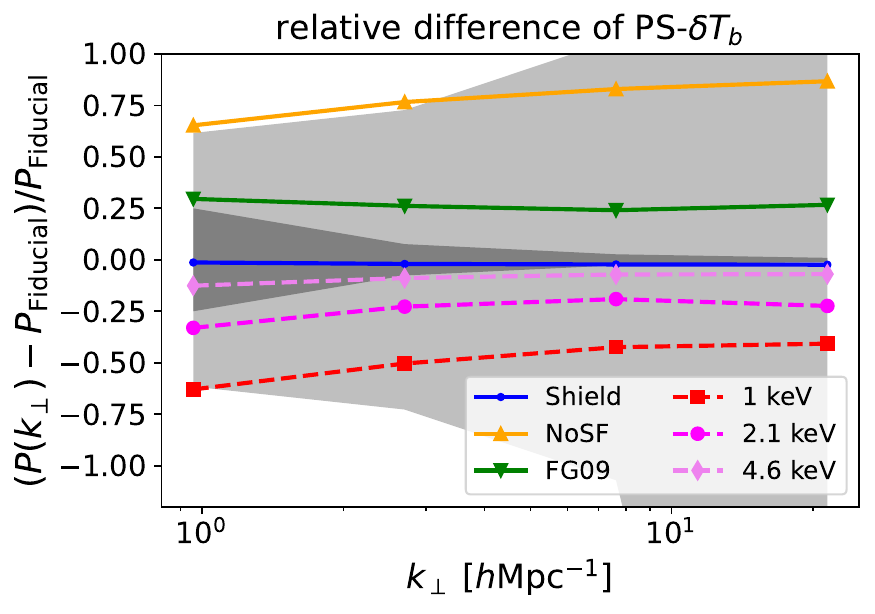}
    \caption{Relative difference between the $\delta T_b$ 2D power spectrum with {\Shield} (blue solid), {\NoSF} (orange solid), {\FG} (green solid), 4.6 keV (violet dashed) or 2.1 keV (red dashed) model and {\Fiducial} CDM model. 
    The shaded region shows $1-\sigma$ error of the cosmic variance (inner shaded region) and $1-\sigma$ error of the cosmic variance $+$ the system noise introduced in Section \ref{ssec:noise_effect} with $t_0 = 1,000$ hours (outer shaded region).
    }
    \label{fig:pk_astro}
\end{figure}

\begin{figure}
    \centering
    \includegraphics[width=\linewidth]{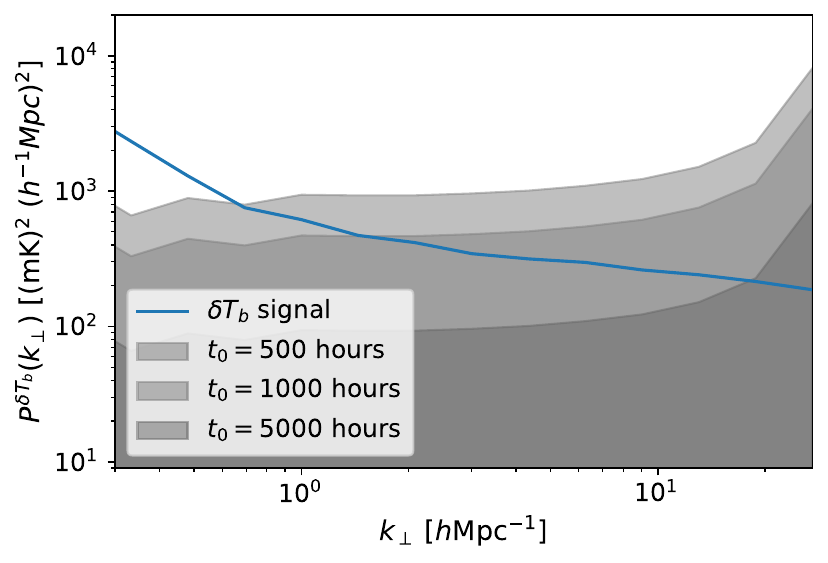}
    \caption{
    This figure shows $P^{\delta T_b} (k_\perp)$ for the {\Fiducial} CDM simulation data (solid line) and the noise power spectra. The outer, middle, and inner shaded regions represent the regions under the noise power spectra with the observation time $t_0=500$, $1000$, and $5000$ hours, respectively.
    }
    
    \label{fig:pk_noise}
\end{figure}

In this subsection, we investigate the effect of the different astrophysical models on the classification. 
To quantify the effect, we replace the CDM test images with the images generated from the simulations of different astrophysical models, i.e., the {\Fiducial} model is replaced with one of {\FG}, {\Shield}, or {\NoSF} models.
In the following, we only consider the analysis of $\delta T_b$ images.

Fig.~\ref{fig:pval_astro} shows the $p$-values of the KS test for different astrophysical models for PS-$\delta T_b$ (left) and CNN-$\delta T_b$ (right).
For PS-$\delta T_b$, the $p$-value 
$< 0.01$ for $m_{\rm DM} \le 2.1$\,keV 
independent of 
the astrophysical models. 
For $m_{\rm DM} \ge 4.6$\,keV, $p$-value is more than 0.1
, so the 2D power spectrum cannot distinguish different dark matter models, and the difference, according to the astrophysical models, is not significant in our power spectrum analysis.

Next, 
the right panel of Fig.~\ref{fig:pval_astro} shows the $p$-value for CNN-$\delta T_b$.
We see that 
for $m_{\rm DM} \le 4.6$ keV, CNN can discriminate the dark matter models regardless of the astrophysical models. 
The $p$-values for \Fiducial, \Shield, and {\NoSF} are comparable, but those for \FG \ show a different behavior.
The $p$-values for the {\FG} models are $< 0.001$ except for $m_{\rm DM} = 4.6$ keV. However, in Table~\ref{tb:astro_AUC}, we see that our CNN cannot correctly classify the images for \FG\ for $m_{\rm DM} > 4.6$ keV. We conclude that the KS test is not valid for the evaluation of our results of the classification between the {\FG} model and for $m_{\rm DM} > 4.6$ keV NCDM models.
Therefore, the results indicate that the astrophysical effects of 
inhomogeneous UV background partly mimics the difference in the density maps between CDM and NCDM. Conversely, for the mass ranges of $m_{\rm DM} < 2.1$ keV, we do not observe the astrophysical effect spoils the classification, and thus, we can conclude that the classification for $m_{\rm DM}<2.1$ keV is robust against the astrophysical models at least within the models we consider in our simulation.

\begin{table}
  \centering
  \begin{tabular}{| c || c | c | c | c |}
    \hline
    Model & 2.1 keV & 4.6 keV & 10 keV & 21 keV \\ \hline
    {\Fiducial} & 1.00 & 1.00 & 0.96 & 0.58 \\
    {\Shield} & 1.00 & 1.00 & 0.96 & 0.58 \\
    {\NoSF} & 1.00 & 1.00 & 0.95 & 0.51 \\
    {\FG} & 1.00 & 0.5 & 0.28 & 0.02 \\
    \hline
  \end{tabular}
  \caption{
  The AUC values for different astrophysical models and dark matter mass models.
  Note that the CNN has been trained assuming the \textit{Fiducial} model.
  The constraints are not affected very much by assuming different models in the cases of the {\Shield} and {\NoSF} models. Still, we find the {\FG} model will be a serious systematic on the classification.
       }
  \label{tb:astro_AUC}
\end{table}

In what follows, we will discuss the effect of astrophysical models on CNN analysis. We quantify the impact using the AUC and confusion matrix.
The confusion matrix is defined as
\begin{equation}
\left(
\begin{array}{crl}
    \displaystyle
    \frac{\mathrm{TP}}{\mathrm{TP}+\mathrm{FN}} &
    \displaystyle
    \frac{\mathrm{FN}}{\mathrm{TP}+\mathrm{FN}} \\
    \\
    \displaystyle
    \frac{\mathrm{FP}}{\mathrm{TN}+\mathrm{FP}} &
    \displaystyle
    \frac{\mathrm{TN}}{\mathrm{TN}+\mathrm{FP}} \\
\end{array}
\right),
\end{equation}
where TP, FN, TN, and FP are 
evaluated
at threshold $t=0.5$. 
The upper left and right elements are 
the rate of the correct and incorrect classification for the NCDM test images and 
lower left and right elements are
the correct and incorrect classification rates for the CDM test images, respectively. 

Figure~\ref{fig:confusion_matrix} 
show the confusion matrix. The left, middle, and right columns correspond to the classification for $m_{\rm DM} = 4.6,10,$ and $21$ keV, respectively, and each row from top to bottom corresponds to the {\Fiducial}, {\Shield}, {\NoSF}, and {\FG} model in Fig.~\ref{fig:confusion_matrix}. We see that there are little differences among the {\Fiducial}, {\Shield}, and {\NoSF} models in the confusion matrix and AUC. However, 
all the CDM images for the {\FG} model are misclassified as NCDM (the lower left element) regardless of the dark matter mass. Accordingly, the AUC value decreases drastically. This is mainly because the {\FG} model has more HI gas than others, affecting the global clustering pattern \citep{Faucher_Gigu_re_2009}. 

We try to understand how discrimination robustness depends on the assumed astrophysical models. First, we consider that the power spectrum conveys most of the information. Thus, we look into the differences in the HI power spectrum in different astrophysical models in comparison with the different dark matter masses.
Fig.~\ref{fig:pk_astro} shows the 
fractional
difference of $P^{\delta T_b} (k_\perp)$
for different astrophysical models or NCDM 
models with different masses, 
compared to the {\Fiducial} CDM power spectrum.
This figure's shaded region in dark grey (inner shaded region) represents the $1 \sigma$ statistical error due to the cosmic variance. 
As shown in this figure, the power spectra for NCDM models are suppressed compared to the {\Fiducial}-CDM model, while the power spectra for other astrophysical models are enhanced.
This partly explains that the dark matter mass can be correctly classified even if we assume different astrophysical models because the effect on the amplitude of the power spectrum is opposite. However, we also see that the {\FG} model also shows the power enhancement, which is supposed to be distinguishable from the dark matter mass model.
Therefore, the power spectrum does not fully explain our results.

In Appendix \ref{sec:hi_halo}, we further explore why the CDM images are misclassified as NCDM when assuming the wrong astrophysical models.
 
\subsection{Effect of System Noise}\label{ssec:noise_effect}

\begin{figure}
    \centering
    \includegraphics[width=\linewidth]{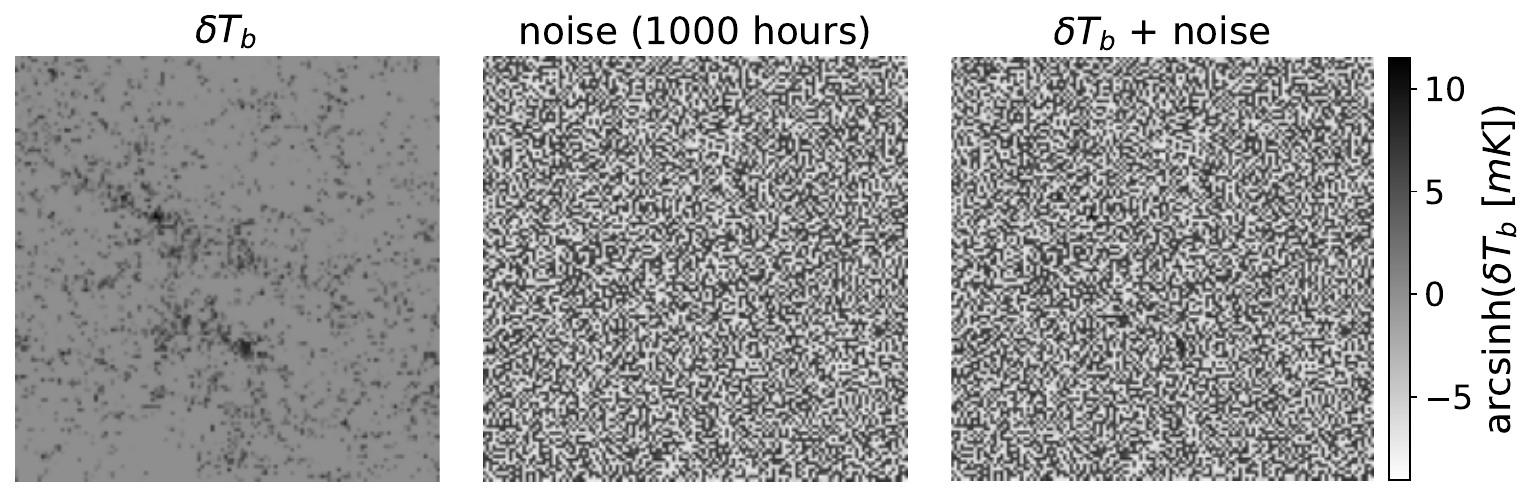}
    \caption{
    Cutout images of $12.5 \times 12.5\ (\mathrm{Mpc}/h)^2$ region of $\delta T_b$ (left), $t_0 = 1,000$ hours noise (middle), and $\delta T_b+$ noise (right) map. 
    The pixel value is $\mathrm{arcsinh} {(\delta T_b\ \mathrm{[mK]})}$.
    }
    \label{fig:noised_image}
\end{figure}

\begin{figure*}
\centering
\begin{tabular}{cc}
  \includegraphics[keepaspectratio,width=0.5\linewidth]{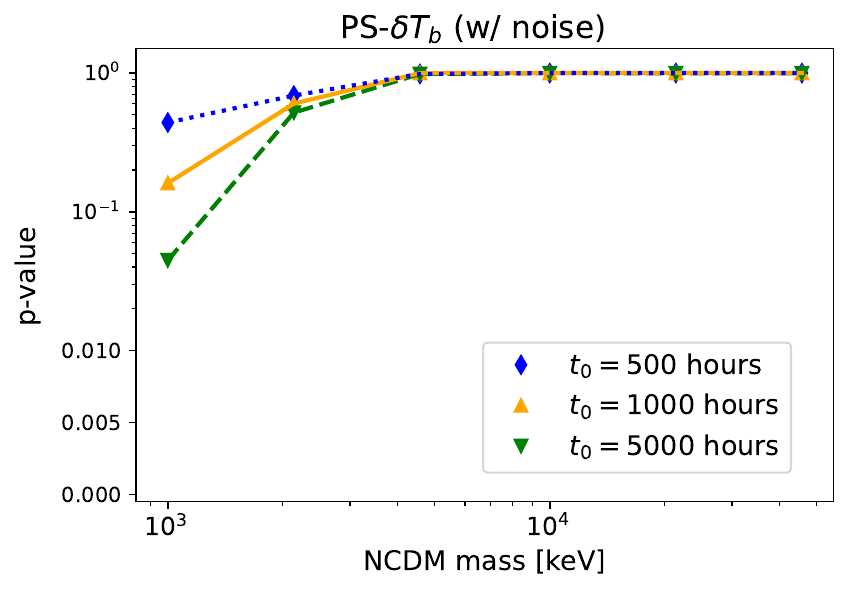} &
  \includegraphics[keepaspectratio,width=0.5\linewidth]{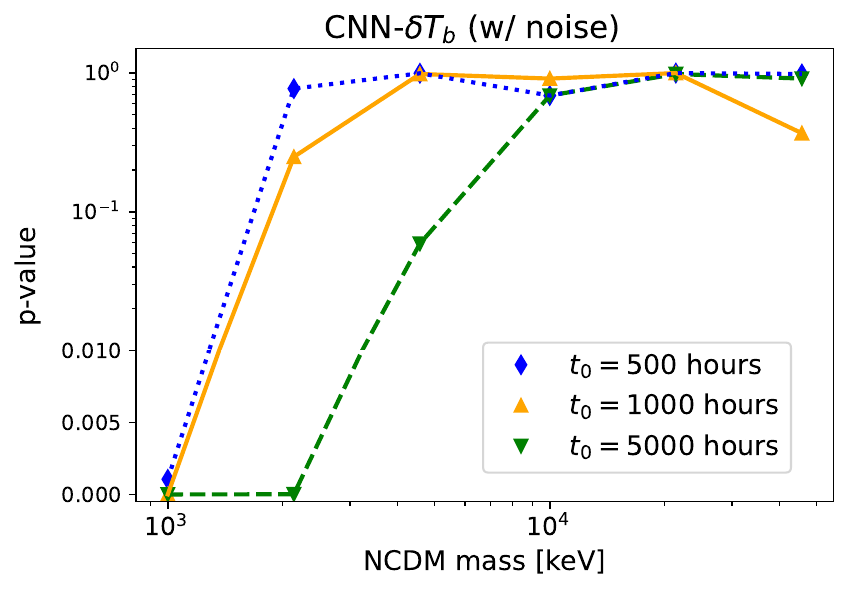}
\end{tabular}
  \caption{Difference in $p$-values of the KS test due to the observation time $t_0=$500 (blue dotted), 1,000 (orange solid), and 5,000 (green dashed) hours for the power spectrum (left panel) and our CNN (right panel). For the same integration time, CNN performs better than the 2D power spectrum. For example, our CNN can distinguish dark matter models for $m_{\rm DM}=$\,1\,keV with $t_0=500$ hours and $m_{\rm DM}=$\,2.1\,keV with $t_0=5,000$ hours with $p$-value $< 0.01$ while the 2D power spectrum cannot distinguish dark matter models $m_{\rm DM}=$\,1\,keV even with $t_0=5,000$ hours.
  \label{fig:pval_noise_include}
  }
\end{figure*}

In this subsection, we 
consider the system noise, particularly assuming the SKA-MID survey, which can observe the 21cm line emission at $0 < z < 3$.

In practice, we should further consider the foreground contamination, such as the Galactic synchrotron emission, the Galactic free-free emission, and the radio emissions from extragalactic sources. These signals are much brighter than the cosmological HI signal. To remove this contamination, various methods are proposed, such as the principal component analysis \citep{2022MNRAS.509.2048S}, generalised morphological component analysis \citep{2020MNRAS.499..304C}, and Gaussian Process Regression \citep{2022MNRAS.510.5872S, 2023MNRAS.524.3724C}. The foreground removal is still an open issue and will require further optimizations. In this work, however, as the most optimistic case, we assume the foreground contamination is completely removed, and let us focus on the potential ability of the CNN applied to the 21cm intensity mapping analysis.

Let us first begin with generating the mock realization of the simulated noise map.
To generate the mock data, including the system temperature noise, 
we first generate the 3-dimensional 
map of the random Gaussian white noise spectrum as a simplistic assumption.
The size of the noise map is $100%^3 
\ h^{-1}\mathrm{Mpc}$ on a side and we define the $1024^3$ grid. The fluctuation of the noise in this map follows the Gaussian distribution of ${\cal N}(0,\sqrt{P_{\rm noise}})$. 
Then, we add this noise map to the simulation box after STEP 1 in Section \ref{ssec:train}. Finally, we generate images following the same procedures in Section \ref{ssec:train} except for the transformation by Eq.~(\ref{eq:asinh_dTb}). 
For the $\delta T_b$ images with the noise, we apply the transformation in the unit of mK instead of nK,
\begin{equation}
    m_T^{\rm obs} = \asinh \left[ \frac{\delta T_b (x) + \mathrm{noise}}{b}\right],
\end{equation}
where we set $b=1 [{\rm mK}]$.
As mentioned in the previous section, we empirically find that the structure at diffuse low-density region has significant information for classifying the dark matter model. Therefore, in the noise-free case, $b=1[{\rm nK}]$ works pretty well.
However,  such low-density regions are highly obscured by the system temperature noise since the typical amplitude of the noise, in our case, is of order mK.
Therefore, we need to focus on the higher density regions where the brightness temperature, $\delta T_b > 1[{\rm mK}]$.

In practice, we find that when $b=1 {\rm mK}$, the CNN can classify the 1 keV NCDM from the CDM model more efficiently than in the case of $b=1 [{\rm nK}]$. For instance, AUC is 0.78 for mK, whereas it decreases to 0.67 for nK case.
We note that the choice of the softening parameter $b$ is still suboptimal, and we can further optimize it; however, we keep $b=1$\,mK here, and do not delve deeper into its exploration for simplicity.

The power spectrum of this Gaussian noise $P_{\rm noise} (\mbf{k}_\perp)$ is written as \citep{Villaescusa-Navarro2015, Geil2010, 2017MNRAS.470.3220W, 2006ApJ...653..815M, 2015ApJ...803...21B}
\begin{equation} \label{eq:pk_noise}
  P_{\rm noise} (\mbf{k}_\perp) = \frac{T^2_{\rm sys}}{2 B t_0} \frac{D^2 \Delta D}{n_{\rm b}(k_\perp D / 2 \pi, \nu)}  \left( \frac{\lambda^2}{A_{\rm e}} \right)^2,
\end{equation}
where
the sensitivity $A_{\rm e} / T_{\rm sys}$ is $\sim 2.3$ at $z\sim 3$ \citep{dewdney2016ska1}, 
 $t_0$ is the total integration time, $D \sim 4400$\,$h^{-1}$Mpc is the comoving distance to the source at $z=3.0$, $\lambda = c(1+z)/\nu_{0}$ is the observed wavelength. $B$ is the bandwidth of the observation and is related to the depth $\Delta D$ for the observation as
 \begin{equation}
     \Delta D \sim \frac{c (1+z)^2 B}{\nu_0 H(z)} .
 \end{equation}
 $n_b (U, \nu)$ is the number density of the baseline, and we use the published data\footnote{https://www.skao.int/en/ska-subarrays}.
%  $n_b (U, \nu)$ is the number density of the baseline given by
% \begin{align}
%     n_b (k_\perp D / 2 \pi, \nu) = &C_b \int^{r_{\rm max}}_0 dr\ 2 \pi r n_a (r) \notag \\
%     &\times \int^{2 \pi}_0 d \phi\ n_a (|\mbf{r} - \lambda \mbf{k}_\perp D / 2 \pi|),
% \end{align}
% where 
% $\nu$ is the observed frequency, $r_{\rm max} = 150$ km is the distance between the centre of antennas and the most distant antenna.
% $n_a$ is the number density of the antenna given by
% \begin{align}
%     n_a (r) = \begin{cases}
%     n_c  & (r \le r_c) \\
%     n_c \left( r_c / r \right)^2 & (r_c < r < r_{\rm max}) \\ 
%     0 & (r \ge r_{\rm max}) \ ,
%     \end{cases}
% \end{align}
% as a function of the distance $r$ from the center of antennas, where $n_c = N_a / (\pi r^2_c + \ln{(r_{\rm max}/r_c)})$ is the antenna number density of the core, $r_c = 2$ km is the radius of the core, $N_a = 197$ is the number of the antenna, and $C_b$ is the normalization factor defined by $(1 / 2 \pi ) \int k_\perp D^2 n_{\rm b}(k_\perp D / 2 \pi) dk_\perp = N_{\rm a} (N_{\rm a} - 1)/2$. In these calculations, we use the value of parameters taken from some pages\footnote{https://www.astron.nl/telescopes/square-kilometre-array/}\footnote{https://www.skao.int/en/science-users/118/ska-telescope-specifications}\footnote{https://indico.skatelescope.org/event/940/contributions/8511/attachments/\\7800/12784/MASUM\_Dish\%20pdf.pdf}.

Note that Eq.~(\ref{eq:pk_noise}) only depends on $k_\perp$, but this noise power spectrum is the 3-dimensional power spectrum, of which the unit is $\mathrm{mK}^2 (\mathrm{Mpc}/h)^3$. 
Therefore, the noise fluctuations follow the Gaussian distribution of ${\cal N}(0,\sqrt{P_{\rm noise}})$ for the direction perpendicular to the line of sight, and are uncorrelated with each other along the line of sight.

We use images, including the noise, to train and test our CNN. 
We assume the integration time $t_0 =$500 hours and hours 1,000 following \citep{Villaescusa-Navarro2015}, and in addition, we assume $t_0=$ 5,000 hours to test the effect of integration time in Eq.~(\ref{eq:pk_noise}), 
where $t_0=1,000$ is often quoted in the literature
\citep[e.g.,][]{Villaescusa-Navarro2015, Crocce2006, Pritchard2015}. 
Figure~\ref{fig:pk_noise} shows the 2D power spectra for the $\delta T_b$ signal and noise. The solid line shows the $P^{\delta T_b} (k_\perp)$ for the {\Fiducial} CDM model, and the outer, middle, and lower shaded regions represent the regions under the noise power spectra with $t_0=500$, $1000$, and $5000$ hours, respectively. These spectra are calculated for the simulation, and the 2D random Gaussian map is generated by projecting the 3D noise map.
Fig.~\ref{fig:pk_astro} shows the relative difference of the 2D power spectrum for different dark matter masses and astrophysical models compared to the error budget, cosmic variance and system temperature noise. We clearly see that the discrimination of the dark matter model becomes challenging in the presence of the system noise. Fig. \ref{fig:noised_image} shows a example of the image including the noise with the signal-only and noise-only image.

As we can see in Fig.~\ref{fig:pval_noise_include}, the 
PS analysis
can distinguish dark matter models for $m_{\rm DM}=$ 1\,keV 
if we have enough integration time, 
$t_0=5,000$ hours 
at more than $2\sigma$,
but for 
observation time $t_0 = 1,000$ hours, we cannot discriminate the dark matter model of $m_{\rm DM}=1$ keV from the CDM.
Even with the noise, the CNN analysis is again superior to the PS analysis. We see that the 
CNN can distinguish dark matter models for $m_{\rm DM}=$ 1\,keV 
even if the integration time is 
$t_0=500$ hours and for $m_{\rm DM}=$ 2.1\,keV NCDM with $t_0=5,000$ hours.
Although the system noise in $\delta T_b$ images degrades the performance of both 
PS and CNN analyses, CNN still provides better performance than the 2D power spectrum. 
The information on the dark matter particle mass for $m_{\rm DM} = 1$\,keV cannot be captured by the power spectrum with $t_0 = 1,000$ hours, hidden below the system noise of SKA-MID, and the CNN can successfully extract it.

In addition, we 
discuss
the effect of the different astrophysical models 
under the existence of
the system noise of SKA. 
The classification with our CNN for $m_{\rm DM} > 4.6$ keV is largely affected by the {\FG} model as shown in Fig.~\ref{fig:pval_astro}.
However, with the 
observation time $t_0=1,000$ hours, the system noise hides the signals
for $m_{\rm DM} \ge 2.1$\,keV as shown in Fig.~\ref{fig:pval_noise_include}. 
Therefore, the difference in the astrophysical model should not be seriously considered given the observational errors in the era of SKA, but it must be the most serious systematic effect in future higher sensitivity observations.

Finally, we consider the effect of the survey area on our results. Our test images covered a $(100  \ h^{-1}\mathrm{Mpc})^2$ area across three redshift slices, corresponding to about 5\,$\mathrm{deg}^2$ of sky at redshift $z=3$. To investigate the impact of the survey area, we derive the $p$-values of the KS test for a limited number of test samples. Specifically, we test the effect of using test images from half of the simulation volume, corresponding to a survey area of 2.5 square degrees. As the survey area increases, the $p$-values decrease for the classification for $m_{\rm DM} \le 4.6$ keV dark matter models. This suggests that increasing the survey area could improve the accuracy of our dark matter mass constraints.

\section{Summary}\label{sec:summary}
 In summary, this paper explores the use of CNN to distinguish between different models of dark matter based on images and 2D power spectra of 21cm brightness temperature distribution. 
 We have shown that the CNN can 
 better distinguish between different dark matter particle masses than the 2D power spectrum. 
 We conduct a suite of hydrodynamic simulations with different dark matter particle masses and generate the images of dark matter distribution and $\delta T_b$ map. In addition, we perform three additional simulations for the CDM model, where the 
astrophysical models
 such as self-shielding of HI gas, star formation, and UV background are different from the {\Fiducial} simulation, following  \citet{Nagamine_2021}. 
 We also injected the system noise from upcoming SKA-MID observation into the $\delta T_b$ images to investigate the effect of noise. 
 
 Firstly, we compare the analysis of dark matter images and $\delta T_b$ images 
 assuming the {\Fiducial} astrophysical model.
 Our results indicate that the direct observable $\delta T_b$ map can constrain the dark matter mass and has comparable classification power to the dark matter image.
 We then compare the performance of our CNN and the 2D power spectrum, finding that our CNN can distinguish dark matter models for $m_{\rm DM} \le 10$ keV, while the 2D power spectrum is only able to distinguish models for $m_{\rm DM} \le 2.1$ keV.
 Therefore, we confirm that the CNN can extract the information not encoded in the 2D power spectrum, which is expected due to the nonlinear evolution of dark matter, which scrambles the Gaussian information at the initial condition, and the nonlinear relation between dark matter and neutral hydrogen.
 
 Secondly, we explore how different
 astrophysical models affect the analysis using power spectrum and CNN.
 To do so, we replace the 
 CDM test images for the {\Fiducial} astrophysical model with those for other astrophysical models.
We find that the power spectrum analysis can distinguish the dark matter models for $m_{\rm DM} \le 2.1$,keV from CDM, regardless of the astrophysical model assumed. 
 The CNN analysis can distinguish dark matter models of  $m_{\rm DM} \le 10$\,keV independent of the assumed astrophysical models except for the {\FG} model.
For the {\FG} model, the classification for $m_{\rm DM} \ge 2.1$\,keV model is highly disturbed.
 
 Thirdly, we investigate the impact of system temperature noise assuming the SKA-MID observation for the $\delta T_b$ map on our classification analysis. We find that the noise significantly degrades the classification performance, but our  CNN can still distinguish the NCDM model with $m_{\rm DM}<1$\,keV from the CDM model with an integration time of  $t_0=500$ hours. With more integration time of $t_0=5000$ hours, this limit can be extended to $m_{\rm DM}<2.1$\,keV. 
     
Finally, we also investigate the effect of the survey area on our analysis. 
Our simulations correspond to a survey area of 5 deg$^2$ at $z=3$, but by scaling the number of test images, we find the probability that the $p$-values for the classification for the $m_{\rm DM} \le 4.6$\,keV dark matter model can be improved.
 
 Our work demonstrates that CNNs have the potential to more effectively constrain the dark matter particle mass than the 2D power spectrum using the $\delta T_b$ map, which can be observed by radio observation like SKA. 
 However, practical observations come with their challenges, such as foreground contamination and the optimal redshift for constraining the dark matter mass. In addition, we can obtain the 3D map of the 21cm and can measure the 3D power spectrum such as a delay power spectrum \citep{2012ApJ...756..165P}. We can use multiple images from various frequencies as inputs of CNN. The information of the 3D 21cm map probably improves the constraints of the dark matter mass.
 We plan to address these challenges in future work.

\section*{Acknowledgements}

We are grateful to Kiyotomo Ichiki, Hironao Miyatake and Shiro Ikeda for fruitful discussions. This work is supported by Japan Science and Technology Agency (JST) AIP Acceleration Research Grant Number JP20317829 and JSPS Kakenhi Grants 18H04350, 21H05454, 21K03625 and 22K21349.
The author (K.M.) would like to thank the “Nagoya University Interdisciplinary Frontier Fellowship” supported by Nagoya University and JST, the establishment of university fellowships towards the creation of science technology innovation, Grant Number JPMJFS2120.
We are grateful to Volker Springel for providing the original version of \texttt{GADGET-3}, on which the \texttt{GADGET3-Osaka} code is based. 
K.N. acknowledges the support from JSPS KAKENHI grants 19H05810, 20H0018, and 22K21349.
K.N. acknowledges the support from the Kavli IPMU, World Premier Research Center Initiative (WPI), where part of this work was conducted. 
Part of the computation is performed on Cray XC50 and GPU cluster at the CfCA in NAOJ, as well as the GPU workstation at Nagoya University.

%%%%%%%%%%%%%%%%%%%%%%%%%%%%%%%%%%%%%%%%%%%%%%%%%%
\section*{Data Availability}

The code for the machine learning used in this work is shared in the GitHub repository, \url{https://github.com/murakoya/IM_ML}



\bibliographystyle{mnras}
\bibliography{bibtex} % if your bibtex file is called example.bib

%\begin{thebibliography}{99}
%\bibitem[\protect\citepauthoryear{Author}{2012}]{Author2012}
%\bibitem[\protect\citepauthoryear{Others}{2013}]{Others2013}
%\end{thebibliography}

%%%%%%%%%%%%%%%%%%%%%%%%%%%%%%%%%%%%%%%%%%%%%%%%%%

%%%%%%%%%%%%%%%%% APPENDICES %%%%%%%%%%%%%%%%%%%%%

\appendix

\section{Property of HI Halo}\label{sec:hi_halo}

\begin{table}
  \begin{tabular}{| c || c | c | c |}
    \hline
    Model & $\mu$ [$h^{-1}$kpc] & $\sigma$ [$h^{-1}$kpc] & Total Number\\ \hline
    {\Fiducial} & 66.2 & 28.2 & 270,931 \\
    {\Shield} & 66.2 & 28.2 & 270,930 \\
    {\NoSF} & 66.2 & 28.2 & 271,092 \\
    {\FG} & 66.2 & 28.2 & 270,936 \\
    \hline
    46 keV & 66.2 & 28.2 & 269,283 \\
    21 keV & 66.2 & 28.5 & 264,444 \\
    10 keV & 66.2 & 28.5 & 245,598 \\
    4.6 keV & 65.2 & 31.9 & 196,314 \\
    2.1 keV & 61.9 & 34.6 & 126,987 \\
    1 keV & 59.9 & 36.7 & 54,755 \\
    \hline
  \end{tabular}
  \caption{Radial size of dark matter halos and the number of halos for different astrophysical and dark matter models. The first four rows are the CDM with different astrophysical models, and the latter six rows are the NCDM models. The second and third column shows the mean ($\mu$) and standard deviation ($\sigma$) of the virial radius of the halo, and the last column shows the total number of the halo in the simulation box. Different astrophysical or dark matter models do not significantly affect the halo size. However, the number of halos for the light dark matter models has decreased.}
  \label{tb:HI_halo_size}
\end{table}

\begin{figure*}
\centering
\begin{tabular}{cc}
  \includegraphics[keepaspectratio,width=\linewidth]{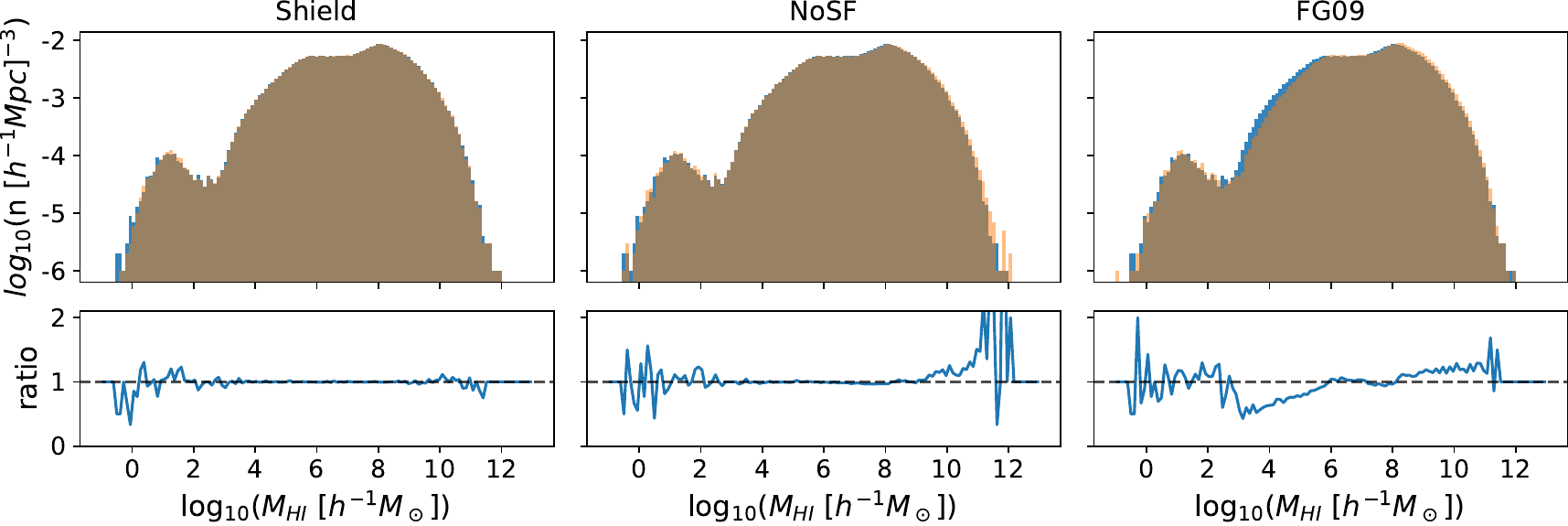} \\
\end{tabular}
  \caption{
  Halo HI mass function (i.e., the comoving number density of halos as a function of their mass) for {\Fiducial} (blue) and other astrophysical models (orange) in the top panels and the ratio of them in the bottom panels. 
  }
  \label{fig:hi_mass_pdf_astro}
\end{figure*}

\begin{figure*}
\centering
\begin{tabular}{cc}
  \includegraphics[keepaspectratio,width=\linewidth]{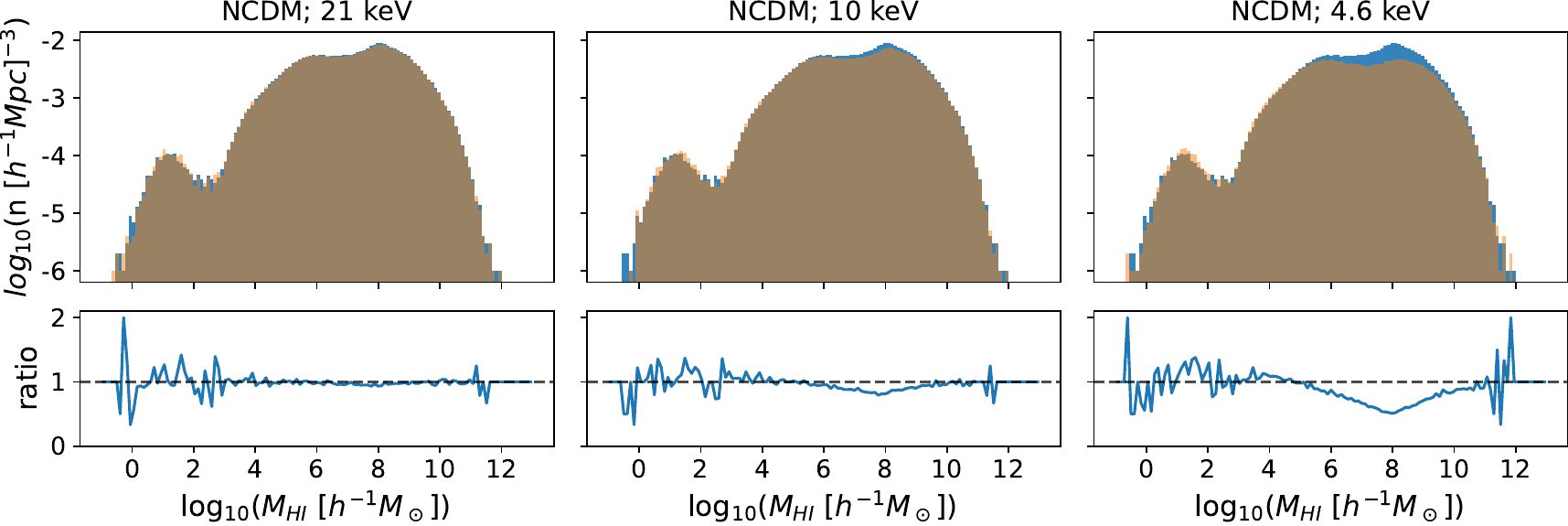} \\
\end{tabular}
  \caption{
  Halo HI mass function for {\Fiducial} (blue) and other NCDM models (orange) in the top panels and the ratio of them in the bottom panels. 
  }
  \label{fig:hi_mass_pdf_ncdm}
\end{figure*}

\begin{figure*}
\centering
\begin{tabular}{cc}  \includegraphics[keepaspectratio,width=0.8\linewidth]{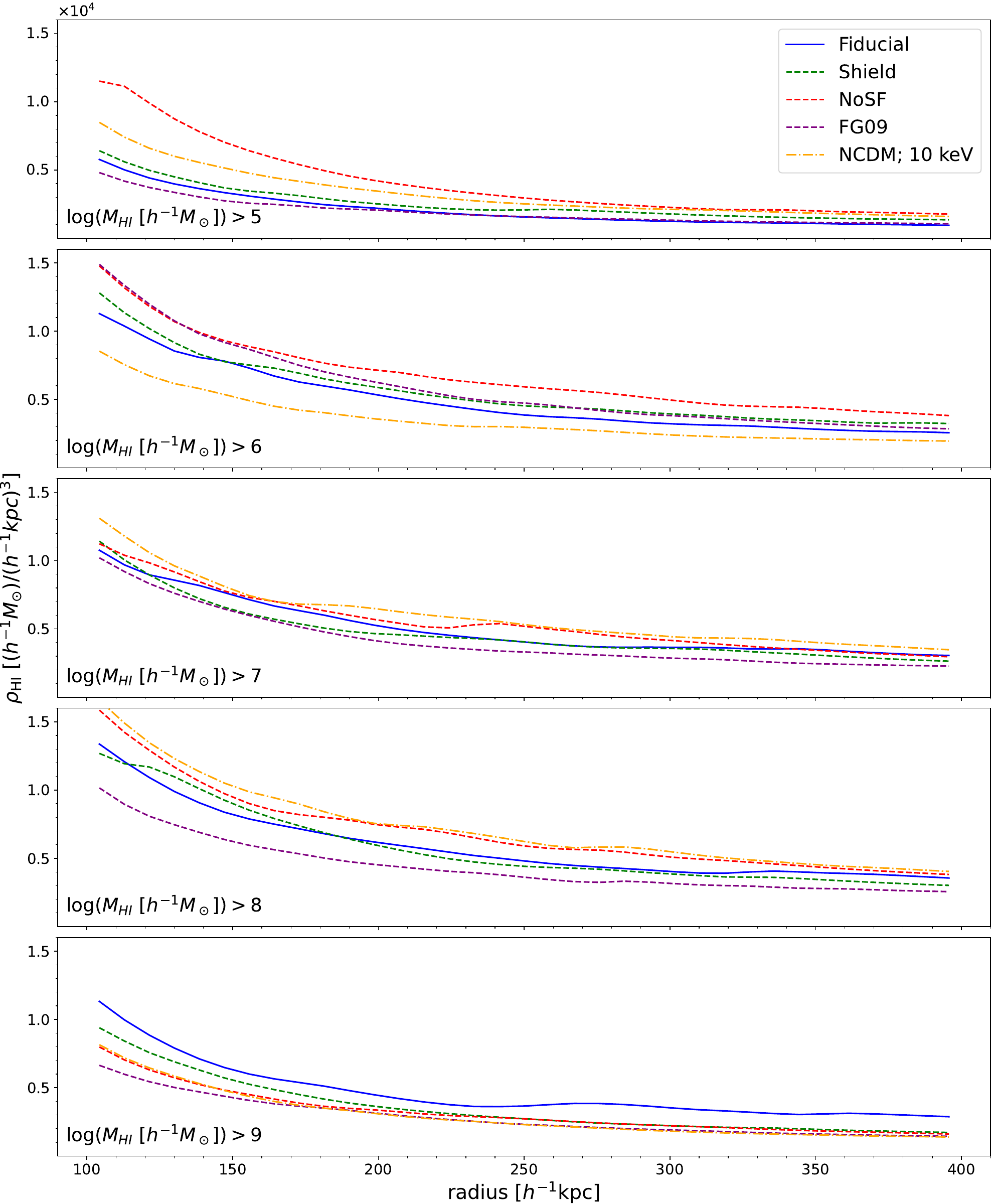}
\end{tabular}
  \caption{Stacked HI density profile as a function of halo-centric radius. We use up to the 3000th halo counted from the lower end in each mass bin of $\log{(M_{\rm HI}\ [h^{-1}M_\odot])} \ge$ 5, 6, 7, 8 and 9 for stacking. Each line corresponds to {\Fiducial} (blue solid), {\Shield} (green dashed), {\NoSF} (red dashed), {\FG} (purple dashed), and 10 keV NCDM (orange dash-dot). The lower limit of abscissa corresponds to the pixel size of the images we use. 
  \label{fig:hi_profile}
  }
\end{figure*}

As discussed in Fig.~\ref{fig:pval_astro}, many of the {\FG} CDM test images is misclassified to the NCDM model. In this appendix, we try to address how this confusion happens by focusing on the fundamental quantities of halos.
We identify the dark matter halo using the ROCKSTAR code \citep{rockstar} and define the HI halo as the group of the HI gas particles within the virial radius of the dark matter halo.

Table~\ref{tb:HI_halo_size} shows the size of the halo, which is defined as the virial radius of the dark matter halo and the total number of halos in the simulation box for each astrophysical and dark matter model. 
We always fix the dark matter model to CDM for the variant run of the astrophysical model, and for the NCDM run, we apply the {\Fiducial} astrophysical model.
We see no significant relation between the number of halos and the assumption of the astrophysical models. In contrast, the total number of halos is smaller when the dark matter mass is smaller, especially for $m_{\rm DM} \le 4.6$ keV. This is because the light dark matter prevents the small-scale clustering due to its velocity dispersion (see Section~\ref{ssec:wdm}), and halos are not formed. On the other hand, the size of the halos is not affected by either the astrophysical models or the dark matter models.

In Fig.~\ref{fig:hi_mass_pdf_astro} and Fig.~\ref{fig:hi_mass_pdf_ncdm}, the panels show the 
HI mass function in comparison with the {\Fiducial} CDM model.
These two figures show the mass function of the astrophysical and NCDM models, respectively.
We 
see that the number of 
halos  
of $M_{\rm HI}> 10^{9} M_{\odot}/h$ 
increases
in the {\NoSF} and {\FG} model compared to the {\Fiducial} model and the 
halos of $M_{\rm HI} \sim 10^8\ h^{-1} M_{\odot}$ decrease for
$m_{\rm DM} \le 10$ keV NCDM models. 
However, these features do not explain the confusion of the model classification because there are no similarities between the HI halo mass function for the NCDM model and variant astrophysical models.

Then, we will see if the halo profile looks similar between the NCDM and variant astrophysical models.
In Fig.~\ref{fig:hi_profile}, the panels show the stacked HI density profile of the halo. 
To compute the stacked HI density profile, we average the HI mass within the dark matter virial radius over the lowest 3000 halos within each mass bin from $10^5$ to $10^9 h^{-1}M_{\odot}$. The {\Fiducial} (blue solid) and {\Shield} (green dashed) runs have relatively similar profiles. In addition,  {\NoSF} (red dashed) and 10 keV NCDM (orange dash-dot) have similar profiles except for halos with $M_{\rm HI} > 10^6\ h^{-1}M_\odot$. However, {\NoSF} model has little effect on the classification in Section~\ref{ssec:astro_effect}. For {\FG} model, its profiles (purple dashed) deviate from {\Fiducial}, especially for massive halos ($M_{\rm HI} > 10^8\ h^{-1}M_\odot$), but they are also different from 10 keV NCDM profile. 
% As we said in Section~\ref{ssec:astro_effect}, 
Our CNN classification is probably based 
not only on features that resemble NCDM but also on features that are not CDM-like.

%%%%%%%%%%%%%%%%%%%%%%%%%%%%%%%%%%%%%%%%%%%%%%%%%%

% Don't change these lines
\bsp	% typesetting comment
\label{lastpage}
\end{document}